\documentclass[%
 reprint,
 amsmath,amssymb,
 aps,
 prl,
 longbibliography,
 lengthcheck,%
]{revtex4-1}

\usepackage{graphicx}% Include figure files
\usepackage{dcolumn}% Align table columns on decimal point
\usepackage{bm}% bold math
\usepackage{hyperref}% add hypertext capabilities
\begin{document}

\preprint{APS/123-QED}

\title{Self-consistent field theory of polarized BEC: dispersion of collective excitation
}% Force line breaks with \\

\author{P. A. Andreev}%
\email{andreevpa@physics.msu.ru}
%[Also at ]
% \altaffiliation{%
\author{L. S. Kuzmenkov}%
\email{lsk@phys.msu.ru} \affiliation{
% Department of General Physics,
 Faculty of Physics, Moscow State
University, Moscow, Russian Federation.}%Lines break automatically or can be forced with \\

%\affiliation{%
% Department of Theoretical Physics, Physics Faculty, Moscow State
%University, Moscow, Russian Federation.}%


%\affiliation{
%\affiliation{
%\affiliation{%


\date{\today}% It is always \today, today,

\begin{abstract}
We suggest the construction of a set of the quantum hydrodynamics
equations for the Bose-Einstein condensate (BEC), where atoms have the
electric dipole moment. The contribution of the dipole-dipole
interactions (DDI) to the Euler equation is obtained. Quantum
equations for the evolution of medium polarization are
derived. Developing mathematical method allows to study
effect of interactions on the evolution of polarization. The
developing method can be applied to various physical systems in
which dynamics is affected by the DDI. Derivation of Gross-Pitaevskii
equation for polarized particles from the quantum hydrodynamics is described. We
showed that the Gross-Pitaevskii equation appears at condition when all
dipoles have the same direction which does not change in time.
Comparison of the equation of the electric dipole evolution with the
equation of the magnetization evolution is described. Dispersion of
the collective excitations in the dipolar BEC, either affected or not affected by
the uniform external electric field, is considered using our
method. We show that the evolution of polarization in the BEC leads to
the formation of a novel type of the collective excitations. Detailed
description of the dispersion of collective excitations is presented.
We also consider the process of wave generation in the polarized BEC
by means of a monoenergetic beam of neutral polarized particles. We
compute the possibilities of the generation of Bogoliubov
and polarization modes by the dipole beam. 
\end{abstract}

%Use showkeys class option if keyword
\maketitle

\section{\label{sec:level1}I. Introduction}

After obtaining the Bose-Einstein condensate (BEC) in experiments
with vapors of alkaline metal atoms, theoretical and experimental
investigation of linear waves and nonlinear structures in the BEC have
been performed. In recent years the interest to the polarized BEC has
been increasing. It is connected with recent experimental progress in
cooling of polarized atoms and molecules. At present day the
BEC with magnetic polarization is realized on atoms $^{52}$Cr. There
are a lot of attempt of experimental obtaining of the electrically
polarized BEC (see reviews ~\cite{Koberle NJP 09}-~\cite{Ni PCCP 09}). For this aim, Bose
molecules having the electric dipole moment have been cooled. Particular interest to the electrically
polarized BEC brought because of large dipole-dipole scattering
length, and thereby because of both the big magnitude and large
distance of interaction in compare with analogous quantities for the
magnetized BEC.

Many processes in quantum systems are determined by the dynamics
and the dispersion of collective excitations (CE) ~\cite{Griffin
book}. The law dispersion of the CE in the degenerate dilute Bose gas was
obtained by Bogoliubov in 1947 ~\cite{Bogoliubov
N.1947,L.P.Pitaevskii RMP 99}. Many authors studied the change of
the Bogoliubov spectrum which arises when the short-range interaction
accounted more carefully ~\cite{Pu PRL 02}-~\cite{Andreev PRA08},
or geometry of the system is complex ~\cite{Stringari PRL 96},
~\cite{Falco PRA 07}, ~\cite{mass dep of waves in BEC}. In
papers ~\cite{Santos PRL 03}-~\cite{Ticknor PRL 11} authors
studied the influence of the electric dipole moment (EDM) dynamics on
the dispersion of the CE in the BEC. The contribution of polarization in the
dispersion law of the Bogoliubov mode was obtained in Ref.s ~\cite{Santos
PRL 03}-~\cite{Ticknor PRL 11}. Instability of the Bogoliubov
spectrum in the 3D dipolar BEC (DBEC) with the repulsive short-range
interaction (SRI) was shown in Ref.s ~\cite{Santos PRL
03}-~\cite{Giovanazzi EPJD 04}. U. R. Fisher ~\cite{Fischer PRA
06R} obtained that the Bogoliubov mode in the 2D DBEC is stable for a
wide range of system parameters. There are also review papers
~\cite{Carr NJP 09}-~\cite{Ni PCCP 09}, where various aspects of
physics of the polarized BEC were considered. In this paper we are
interested in possibility of the polarization wave existence in the
DBEC, i.e. in existence of new type of the CE. Dynamics of polarization is
interesting not for the quantum gases only, but in the solid state physics
and the physics of low-dimensional systems too. The polarization waves
in the low dimensional and the multy-layer systems of conductors,
dielectrics, and semiconductors are considered in the papers ~\cite{Andreev PRB
11,Qiuzi Li PRB 11}. Analogously, in the BEC of molecules having the EDM we
expect the existence of a polarization wave along with the
Bogoliubov mode.

For investigation of the BEC, in system of particles with the EDM or the
magnetic moments, various theoretical methods are used. One of the
ways of theoretical description of the polarized BEC is the
generalization of well-known Gross-Pitaevskii (GP) equation
~\cite{Yi PRA 00}-~\cite{Goral PRA 02}. In this way next papers
were made: for the spinor BEC ~\cite{Szankowski PRL 10,Cherng PRL
09,Lahaye RPP 09}, for the effect of magnetic moment on the BEC
evolution ~\cite{Lahaye RPP 09}-~\cite{Lahaye Nature} and for the
influence of electrical polarization of atoms on the dynamic
processes in the BEC ~\cite{Lahaye RPP 09}-~\cite{polarized BEC first
step}. Dipolar fermions also attract a lot of attention ~\cite{Lu
11 arxiv}, ~\cite{Gadsbolle 11 arxiv}.

The same generalization of the GP equation is usually used for the magnetic
moments and the electric dipoles:
$$\imath\hbar\partial_{t}\Phi(\textbf{r},t)=\biggl(-\frac{\hbar^{2}\nabla^{2}}{2m}+\mu(\textbf{r},t)+V_{ext}(\textbf{r},t)+g\mid\Phi(\textbf{r},t)\mid^{2}$$
\begin{equation}\label{di BEC GP eq for introduction}+d^{2}\int d\textbf{r}'
\frac{1-3\cos^{2}\theta'}{|\textbf{r}-\textbf{r}'|^{3}}\mid\Phi(\textbf{r}',t)\mid^{2}\biggr)\Phi(\textbf{r},t)
.\end{equation}
In equation (\ref{di BEC GP eq for introduction}) following designation are used: $\Phi(\textbf{r},t)$ is the macroscopic wave function, $\mu$ is the chemical potential, $V_{ext}$ is the potential of external field, $g$ is the constant of short-range interaction, $d$ is the dipole electric moment of single atom, $m$ is the mass of particles and $\hbar$ is the Planck constant divided by $2\pi$.

Along with the GP equation the method of quantum hydrodynamics (QHD) is also used. The method of QHD ~\cite{Andreev PRA08}, ~\cite{MaksimovTMP 1999}, ~\cite{Andreev arxiv 12 01} was used for
investigations of various physical systems, particularly for the BEC
~\cite{Andreev PRA08}, ~\cite{Andreev arxiv ThPart} and the polarization waves in conductors and
semiconductors ~\cite{Andreev PRB 11}. Therefore, the method of QHD could be used to consider the
possibility of arising polarization waves in the BEC and calculate
of the dispersion law of these waves.

At interpretation of the DDI in the quantum gases (QG) used the notion of the
scattering length (SL) and the first Born approximation (FBA),
analogously to the SRI described by the fourth term in the right-hand side
of equation (\ref{di BEC GP eq for introduction}). Therefore, the
problem reduces to the process of scattering. Different
approximations based on scattering process was analyzed in Ref.
~\cite{Wang NJP 08}. Particularly, where were considered condition
of the FBA using and presented generalization of the GP equation
for the scattering of polarized atoms beyond the FBA.

\begin{figure}
\includegraphics[width=8cm,angle=0]{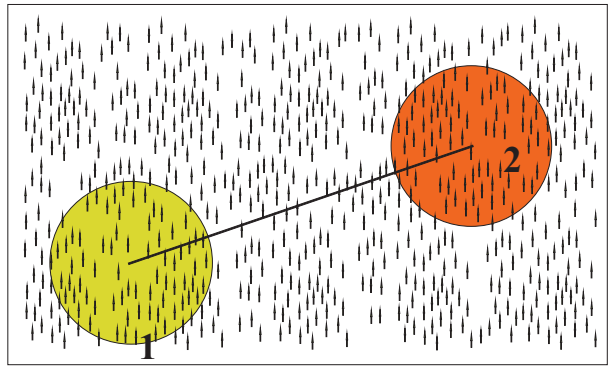}
\caption{\label{dipBECint01} (Color online) The figure presents the picture of the
self-consistent interaction between dipoles. Total polarization of
region 2 (red) interacts with total polarization of region 1
(green). Changing the extreme point of a radius vector (or
shifting the region 2) we can scan whole space. In this way we
obtain action of external dipoles on region 1. Changing position
of region 1 and repeating described operation we obtain action of
surrounding dipoles on each region of space. This is a picture of the
self-consistent interaction in fixed moment of time and this
picture governs an evolution of polarization in system. This
picture of interaction is typical for the classic physics, where we
need to obtain smooth functions describing collective motion. For
this aim is necessary to average the collective variables at the physically infinitesimal volume
(sketched circle). In quantum mechanics, where the particle concentration,
polarization of medium, etc, are defined via wave function and we can consider
described picture on interaction of separate particle instead of
space regions.}
\end{figure}

Considering the QHD we do not formulate the scattering problem.
We do not need to limit ourselves to a scattering process,
because we are interested in interaction between atoms including
interaction between several atoms at the same time. This statement is
especially important for the polarized atoms because of the long-range
interaction between dipoles. Each act of interaction between dipoles
could be considered as the scattering process for system of particles
with a low concentration. This approximation is
possible as for the neutral polarized particles as for the charged
particles. The kinetic theory for the last case was developed by
L. D. Landau in 1936 ~\cite{Landau v10}. Landau constructed
collision term for the Coulomb particles, using analogy with the
Boltzmann equation. The Landau collision integral depends on a scattering
cross section. In this connection he considered the problem of the
scattering of charged particles on small angles, that is suitable
for rare systems. Strictly speaking, for particles with the long-range
interaction the conception of the self-consistent field is more
suitable. This conception was suggested by A. A. Vlasov in 1938,
for system of the charged particles ~\cite{Vlasov JETP 38}. In fact,
group of particles moves in the field of their neighbors, but
state of motion of the neighbors depends on motion of this group of
particles as well (This picture of interaction is described on Fig. \ref{dipBECint01}). According to this idea
Vlasov proposed a kinetic equation for the charged particles called the
Vlasov equation. Actually the Landau collision integral and the
Vlasov equation are the two opposite branches of a hypothetical
general theory. In this paper we follow to the Vlasov approximation
and derive the self-consistent field theory for the dipole-dipole
interaction in the quantum gases. From this point of view, there is no
need to consider the first Born approximation or scattering
process. In Appendix B to this paper we present derivation
of the GP equation for polarized particles from the QHD equations. We did
it at the condition that dipoles are parallel each other and to an
external field, they directions do not oscillate or vibrate
around an equilibrium direction.

Other methods were also applied for investigation of the polarized
BEC and other systems of particles having the EDM, along with the GP
equation. A hydrodynamic formulation of the Hartree-Fock theory
for particles with the significant EDM is considered in Ref. ~\cite{Lima
PRA 10}. The Euler-type equation was
derived in paper ~\cite{Lima PRA 10}, from the evolution of the density matrix. The EDM
dynamics in dimer Mott insulators causes the rise of the
low-frequency mode ~\cite{Gomi PRB 10}. The spectrum of single-particle excitations and
long-wave-length collective modes (zero sound) in the normal phase
were obtained ~\cite{Sieberer PRA 11} in the Hartree-Fock
approximation, which treats direct and exchange interactions on an
equal footing. Minimal coupling model for
description of spatial polarization changing on the BEC properties is
constructed in Ref. ~\cite{Wilson arxiv 11}. The density functional method was used for the polarized
quantum gas studying in Ref. ~\cite{Fang
PRA 11}. The Hubbard model for the polarized BEC is used
too ~\cite{He PRA 11}. A two-body quantum problem for the polarized
molecules is analyzed in ~\cite{Quemener PRA 11}. The existence of
states with spontaneous interlayer coherence has been predicted in
~\cite{Lutchyn PRA 10} in systems of the polar molecules.
Calculations, in ~\cite{Lutchyn PRA 10}, were based upon the
secondary quantization approach, where the Hamiltonian accounts
for the molecular rotation and the dipole-dipole interactions.
Superfluidity anisotropy of polarized fermion systems was shown in
~\cite{Liao PRA 10} and their thermodynamic and correlation
properties were investigated ~\cite{Baillie PRA 10}. The effect of the
EDM on a system of cooled neutral atoms which are used for the quantum
computing and quantum memory devices was analyzed in Ref.
~\cite{Gillen-Christandl PRA 11}.

The characteristic property of the BEC in a system of excitons
inducing in semiconductors ~\cite{Deng RMP 10} is a significant
value of the exciton EDM. This leads to strong interaction of
excitons with an external electric field and emergence of the
collective DDI in exciton systems. The QHD method may be applied to
such systems along with quantum kinetics based on the
nonequilibrium Green functions ~\cite{Haug book 08} or the density
matrix equations ~\cite{Kuhn book 97}.

Notable success has been reached in the Bose condensation of dense
gases ~\cite{Vogl Nature 09,Sheik-Bahae NP 09}. Consequently, we
need the detailed account of the short range interaction for
investigation of the EDM dynamics. In this work we account the
SRI up to the third order on interaction
radius (TOIR) ~\cite{Andreev PRA08}. This approximation leads to
nonlocality of the SRI ~\cite{Andreev PRA08}, ~\cite{Rosanov PL A 02}-~\cite{Andreev arxiv 11 1}.

Electrically polarized BEC can interact with the beam of charged
and polarized particles by means of the charge-dipole and the
dipole-dipole interaction. These interactions lead to transfer of
energy from the beam to medium and, consequently, to generation of
waves. In the plasma physics, the effects of generation of waves by electron ~\cite{Bret PRE 04} or magnetized neutron
~\cite{Andreev PIERS 2011,Andreev arxiv MM}  beam are well-known.
When we use term beam we also mean stream of particles excited in
the considered system, along with an external beam of particles
passing through the system. In presented paper we consider similar
effect in the polarized BEC.

\begin{figure}
\includegraphics[width=8cm,angle=0]{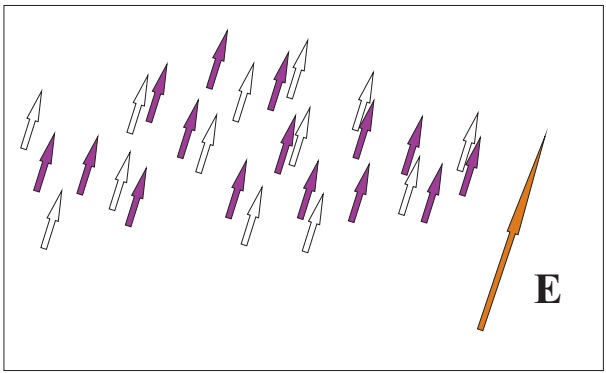}
\caption{\label{dipBECM01} (Color online) The figure shows motion of dipoles
without changing of direction of dipoles during motion. Dipoles
stand at temperature equal to zero $T=0$ and in external uniform
static electric field $\textbf{E}$. This picture describes
particles in two different moments of time. White dipoles presents
system in the first moment of time,
and violet dipoles are present system in the next moment.}
\end{figure}

\begin{figure}
\includegraphics[width=8cm,angle=0]{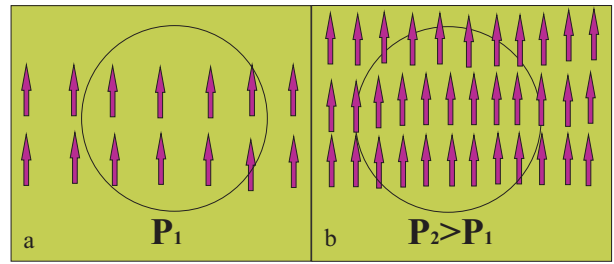}
\caption{\label{dipBECM03} (Color online) The figure presents the change of
polarization density at movement of dipoles in the absence of
dipoles direction evolution.}
\end{figure}

At our treatment we derive and use the QHD equations for polarized
particles. Set of the QHD equations consists of the continuity
equation, the momentum balance equation (the Euler equation), the equation of polarization
evolution and equation of polarization current evolution. We present
first principles derivation of this equation. For this purpose we
use the many-particle Schr\"{o}dinger equation. We
analytically calculate the dispersion properties of the DBEC.
We show that the dynamics of the EDM leads to existence of new
branch in the dispersion law. Consequently, there are the waves of
polarization in the DBEC, along with the Bogoliubov mode. We
obtain contribution of the DDI to the dispersion of the Bogoliubov
mode. Then, we consider the process of wave
generation in the DBEC by means of a monoenergetic beam of
neutral polarized particles. We suggest new method of generation
of both the Bogoliubov mode and the
polarization wave. This paper is an extension and result of
processing of our previous paper ~\cite{Andreev arxiv Pol}, where
a brief description of the same results and ideas were
presented. Some ideas were also described in Ref. ~\cite{Andreev RPJ 12}.

This paper is organized as follows. We introduce the model
Hamiltonian in Sec. II, and present the  momentum balance
equation. Further, in Sec. II, we derive equations
of the polarization evolution and obtain the influence of the interactions on
the evolution of polarization. In Sec. III  we calculate the
dispersion dependence of the CE in the DBEC. The polarization evolution is
taken into account. We obtain the contribution of polarization in
dispersion of the Bogoliubov mode and show the existence of new wave
solution. We consider high frequency excitation appearing at large
equilibrium polarizations. We also present calculations for the
wave of polarization at constant concentration. In Sec. IV we
study the wave generation in the polarized BEC by means of the neutral
polarized particle beam. In Sec. V we present the brief summary of
our results. App. A. contains detail of derivation of equations of
polarization evolution. In App. B. derivation of the GP equation for
polarized particles is presented. In App. C. we consider
distinction of evolution of spinning particle from the particles
having EDM.

\begin{figure}
\includegraphics[width=8cm,angle=0]{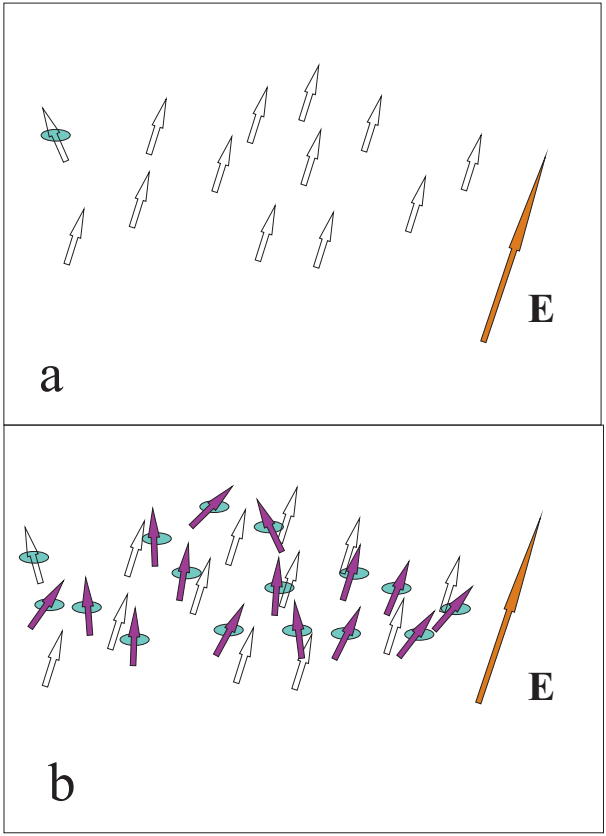}
\caption{\label{dipBECM02} (Color online) The figure describes how the polarization direction perturbation propagates through the system of
particles. We consider system of dipoles at
$T=0$ and being in an external uniform electric field. Upper picture
shows that if at the first moment there is a perturbation of the EDM,
in next moment, this perturbation passes to several neighbors. At
lower picture, particles in two different moments of time are described.
White dipoles presents system in the first moment of time, when one dipole is perturbed
only, but violet dipoles present system in the
next moment.}
\end{figure}

\section{ The model}

\begin{figure}
\includegraphics[width=8cm,angle=0]{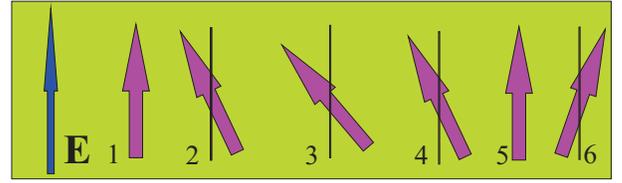}
\caption{\label{dipBECM04} (Color online) The figure presents the
oscillation of the EDM near of the equilibrium position which defined
by the external electric field $\textbf{E}$. Indexes 1-6 present the
different moments of time $t_{1}<t_{2}<...<t_{6}$. This figure
describes mechanism of the polarization change due to oscillations
of the EDM of each particle in the system. This mechanism might reveal
along with rotation of the EDM. Incoherent motion of the EDMs which is
combination of rotation and oscillation is presented on Fig.
\ref{dipBECM02}(b).}
\end{figure}

The term describing DDI in the GP equation, for both the EDM and the magnetic
moment, is interpreted via the dipole scattering length (DSL) $a_{dd}$.
The GP equation does not contain information about evolution of the dipole
direction. Changing of the dipole
direction appears due to the interaction. If we have an external static
uniform electric field, at close to zero temperatures, all dipoles
will be directed parallel to the external field. At low temperatures,
there is no temperature disorientation of dipoles and we have
system of parallel dipoles, as it is shown on Fig.
\ref{dipBECM01}. In such systems, there are two type of dipoles
motion. The first one is a motion, where the direction of dipoles
preserves (see Fig.s \ref{dipBECM01}, \ref{dipBECM03}). The
DDI  leads to anisotropy of interaction in the BEC. At
such form of motion the polarization is changed accordingly to the particle
concentration $\textbf{P}(\textbf{r},t)=\textbf{d}
n(\textbf{r},t)$, where $\textbf{d}$ is the EDM of single particle,
this motion is depicted on the Fig. \ref{dipBECM03}. The second type of motion is a motion of
dipoles including change of dipole direction described on the Fig.s
\ref{dipBECM02}, \ref{dipBECM04}. Perturbation of dipole
direction can be caused by both external influence and local
fluctuation of direction of other dipoles produced by the DDI. In consequence
of the DDI, a perturbation will propagate through the system.
Particularly, this propagation might have wave nature. We need to know how polarization of
system evolves in different point of space during the time for
description of such motion.
Therefore we need to derive equations describing the evolution of
polarization field and accounting influence of interaction on this
evolution.

\begin{figure}
\includegraphics[width=8cm,angle=0]{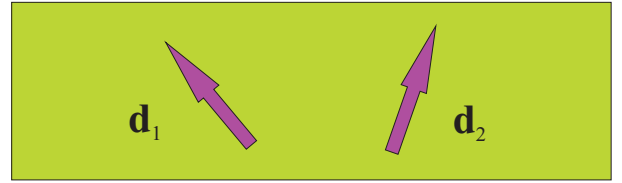}
\caption{\label{dipBECint02} (Color online) The figure presents
the two dipoles with the EDM turned in different directions. At
propagation of perturbation through the system relative directions
of dipoles is changed. This lied to change of force of interaction
in concordance with potential of the DDI (\ref{di BEC Hamiltonian}).
In Appendix B we shown that the GP equation describes motion for
parallel dipoles only.}
\end{figure}

In this case dipoles in different
areas of space have different directions Fig. \ref{dipBECint02} and it reflects on
interaction via angle dependence under integral in the GP equation (\ref{di BEC GP eq for introduction}).
Still, this equation does not contain information about propagation of
dipoles direction perturbation.

At studying we pay attention for the SRI and DDI. We also
consider the action of external fields on particles. We consider the action of
external electric field on dipoles in explicit form and
the external field describing trap presented in general form by
trapping potential $V_{trap}(\textbf{r}_{i},t)$.

We used quasi-static approximation for the DDI. The quasi-static
approximation means that we consider interaction between
dipoles at fixed moment of time as they are motionless. The DDI leads
to change of state of motion, and, in particularly, to variation
of dipole direction. Then, we talk about change of direction we mean
both the variation during precession around of rotation axis and
deviation of rotation axis direction. Described approximation is
an analog of the Coulomb interaction, where interaction between moving
charges approximately described by means of interaction potential
of motionless charges, this approximation usually used for
description of non-relativistic plasmas ~\cite{Akhiezer}.

Explicit form of the Hamiltonian of considering system in the
quasi-static approximation is
$$\hat{H}=\sum_{i}\Biggl(\frac{1}{2m_{i}}\hat{\textbf{p}}_{i}^{2}-\textbf{d}_{i}\textbf{E}_{i,ext}+V_{trap}(\textbf{r}_{i},t)\Biggr)$$
\begin{equation}\label{di BEC Hamiltonian}+\frac{1}{2}\sum_{i,j\neq i}\Biggl(U_{ij}-d_{i}^{\alpha}d_{j}^{\beta}G_{ij}^{\alpha\beta}\Biggr).\end{equation}
The first term in the Hamiltonian is the operator of the kinetic energy. The second
term presents the interaction between the dipole moment
$d_{i}^{\alpha}$ and the external electrical field. The subsequent
terms present the short-range $U_{ij}$ and the dipole-dipole
interactions between particles, respectively. The Green function
of the dipole-dipole interaction reads as
$G_{ij}^{\alpha\beta}=\nabla^{\alpha}_{i}\nabla^{\beta}_{i}(1/r_{ij})$.

For investigation of the CE dynamics in the polarized BEC we derive
set of the QHD equations. This system of equations consists of the
continuity equation, the Euler equation and, for the case of
polarized particles, the equations of polarization evolution and
equations of field (the Maxwell equations). The system of equations is derived by methods
described in Ref. ~\cite{Andreev PRA08}.

The first equation of the QHD equation set is the continuity
equation
\begin{equation}\label{di BEC cont eq}\partial_{t}n(\textbf{r},t)+\partial^{\alpha}(n(\textbf{r},t)v^{\alpha}(\textbf{r},t))=0.\end{equation}
The momentum balance equation for the polarized BEC appears as
$$mn(\textbf{r},t)(\partial_{t}+\textbf{v}\nabla)v^{\alpha}(\textbf{r},t)+\partial_{\beta}p^{\alpha\beta}(\textbf{r},t)$$
$$-\frac{\hbar^{2}}{4m}\partial^{\alpha}\triangle
n(\textbf{r},t)+\frac{\hbar^{2}}{4m}\partial^{\beta}\Biggl(\frac{\partial^{\alpha}n(\textbf{r},t)\cdot\partial^{\beta}n(\textbf{r},t)}{n(\textbf{r},t)}\Biggr)
$$
$$=\Upsilon n(\textbf{r},t)\partial^{\alpha}n(\textbf{r},t)+\frac{1}{2}\Upsilon_{2}\partial^{\alpha}\triangle n^{2}(\textbf{r},t)$$
\begin{equation}\label{di BEC bal imp eq short}+P^{\beta}(\textbf{r},t)\partial^{\alpha}E^{\beta}(\textbf{r},t),
\end{equation}
where
\begin{equation}\label{di BEC Upsilon} \Upsilon=\frac{4\pi}{3}\int
dr(r)^{3}\frac{\partial U(r)}{\partial r},
\end{equation}
and
 \begin{equation}\label{di BEC Upsilon2}\Upsilon_{2}\equiv\frac{\pi}{30}\int dr
(r)^{5}\frac{\partial U(r)}{\partial r}.\end{equation}
In equation (\ref{di BEC bal imp eq short})  we defined a
parameter $\Upsilon_{2}$ as (\ref{di BEC Upsilon2}). This
definition differs from the one in the paper ~\cite{Andreev PRA08}.
Terms proportional to $\hbar^{2}$ appear as a result of using of
quantum kinematics. The first two terms in the right-hand side of the
equation (\ref{di BEC bal imp eq short}) are the first terms of
expansion of the quantum stress tensor. They occur because of
taking into account of the SRI potential $U_{ij}$. The interaction
potential $U_{ij}$ determines the macroscopic interaction
constants $\Upsilon$ and $\Upsilon_{2}$. The last two terms in
the equation (\ref{di BEC bal imp eq short}) describe force fields
that affect the dipole moment in a unit of volume as the effect of
the external electrical field and the field produced by other
dipoles, respectively. The last term is written using the
self-consistent field approximation ~\cite{MaksimovTMP 1999}, ~\cite{Andreev PRB 11}.
$p^{\alpha\beta}(\textbf{r},t)$ is the kinetic
pressure tensor, which depends on the thermal velocities of particles and does
not contribute into the BEC dynamics at temperatures near zero.

The first order of the interaction radius interaction constant
for dilute gases has the form
\begin{equation}\label{di BEC Upsilon via scattering length}\Upsilon=-\frac{4\pi\hbar^{2}a}{m},\end{equation}
where $a$ is the scattering length (SL) ~\cite{L.P.Pitaevskii RMP
99,Andreev PRA08}. The value $\Upsilon_{2}$ may be expressed
approximately as
\begin{equation}\label{di BEC Upsilon 2 connection of Up}\Upsilon_{2}=-\frac{\theta a^{2}\Upsilon}{8},\end{equation}
where $\theta$ is a constant positive value about 1, which depends
on the interatomic interaction potential. Finally, $\Upsilon_{2}$
takes the form
\begin{equation}\label{di BEC Upsilon 2 approx}\Upsilon_{2}=-\frac{\pi\theta\hbar^{2}a^{3}}{2m}.\end{equation}

We have also equations of field
\begin{equation}\label{di BEC field eq}\nabla\textbf{E}(\textbf{r},t)=-4\pi \nabla\textbf{P}(\textbf{r},t),\end{equation}
and
\begin{equation}\label{di BEC field eq rot}\nabla\times\textbf{E}(\textbf{r},t)=0,\end{equation}
which are the Maxwell equation.

In the case of particles having no dipole moment, the
continuity equation and the momentum balance equation form a
closed set of equations. When the dipole moment is taken into
account in the momentum balance equation, a new physical value
emerges, it is the polarization vector field $P^{\alpha}(\textbf{r},t)$.
Thus, the set of equations is not closed.

Next equation we need for studying of the CE
dispersion is the polarization evolution equation
\begin{equation}\label{di BEC eq polarization}\partial_{t}P^{\alpha}(\textbf{r},t)+\partial^{\beta}R^{\alpha\beta}(\textbf{r},t)=0,\end{equation}
where $R^{\alpha\beta}(\textbf{r},t)$ is the current of polarization.

Equation (\ref{di BEC eq polarization}) does not contain
information about the effect of the interaction on the
polarization evolution. An evolution equation of
$R^{\alpha\beta}(\textbf{r},t)$ can be obtained by analogy with
the equations derived above. Method of the equations derivation is
described in Appendix A. Using the self-consistent field approximation
for the dipole-dipole interaction we obtain an equation for the
polarization current $R^{\alpha\beta}(\textbf{r},t)$ evolution
$$\partial_{t}R^{\alpha\beta}(\textbf{r},t)+\partial^{\gamma}\biggl(R^{\alpha\beta}(\textbf{r},t)v^{\gamma}(\textbf{r},t)$$
$$+R^{\alpha\gamma}(\textbf{r},t)v^{\beta}(\textbf{r},t)-P^{\alpha}(\textbf{r},t)v^{\beta}(\textbf{r},t)v^{\gamma}(\textbf{r},t)\biggr)$$
$$+\frac{1}{m}\partial^{\gamma}r^{\alpha\beta\gamma}(\textbf{r},t)-\frac{\hbar^{2}}{4m^{2}}\partial_{\beta}\triangle P^{\alpha}(\textbf{r},t)$$
$$+\frac{\hbar^{2}}{8m^{2}}\partial^{\gamma}\biggl(\frac{\partial_{\beta}P^{\alpha}(\textbf{r},t)\partial_{\gamma}n(\textbf{r},t)}{n(\textbf{r},t)}+\frac{\partial_{\gamma}P^{\alpha}(\textbf{r},t)\partial_{\beta}n(\textbf{r},t)}{n(\textbf{r},t)}\biggr)
$$
$$=\frac{1}{m}\Upsilon\partial^{\beta}\biggl(n(\textbf{r},t)P^{\alpha}(\textbf{r},t)\biggr)$$
\begin{equation}\label{di BEC eq for pol current gen selfconsist
appr}+\frac{\sigma}{m}\frac{P^{\alpha}(\textbf{r},t)P^{\gamma}(\textbf{r},t)}{n(\textbf{r},t)}\partial^{\beta}E^{\gamma}(\textbf{r},t),\end{equation}
where  $r^{\alpha\beta\gamma}(\textbf{r},t)$ represents the
contribution of thermal movement of polarized particles into the
dynamics of $R^{\alpha\beta}(\textbf{r},t)$. As we deal with the BEC,
the contribution of $r^{\alpha\beta\gamma}(\textbf{r},t)$
may be neglected. The last term in the formula (\ref{di BEC eq for
pol current gen selfconsist appr}) includes both external
electrical field and the self-consistent field created by particle dipoles.
This term contains numerical constant $\sigma$.
Approximations used in formula
(\ref{di BEC eq for pol current gen selfconsist appr}) are described in Appendix A.
The first term in the right-hand side of equation (\ref{di BEC eq for pol
current gen selfconsist appr}) describes the short-range
interaction.

We can see various interactions are included in the equations
(\ref{di BEC bal imp eq short}) and (\ref{di BEC eq for pol
current gen selfconsist appr}) additively. At short distances
among particles the SRI and the dipole-dipole interaction act together. At
large distances dipole-dipole interaction remains only.

We can see from equation (\ref{di BEC eq for pol current gen selfconsist appr})
that change of the polarization arises from both the
dipole-dipole interaction and the short range interaction. The SRI
among particle leads to displacement of particles. Consequently,
as particles have the EDM, there is motion of the EDM. Therefore, there is evolution of
$R^{\alpha\beta}(\textbf{r},t)$.

The terms proportional to $\hbar^{2}$ have the quantum
origin. They are analogs of the Bohm quantum potential in the
momentum balance equilibrium.

In the Appendix C we consider the difference between equations of
electrical and magnetic polarization evolution.
The QHD of spinning particles  with spin-orbit interaction were
developed in Ref. ~\cite{Andreev arxiv MM} for system of charged particles, however, obtained
where force field can be used for neutral quantum
gases too.

\section{Collective excitations in the polarized BEC}

We can analyze the linear dynamics of the collective excitations in the
polarized three dimensional BEC using the QHD equations (\ref{di BEC cont eq}),
(\ref{di BEC bal imp eq short}), (\ref{di BEC field eq}), (\ref{di
BEC eq polarization}) and (\ref{di BEC eq for pol current gen
selfconsist appr}). Let's assume that the system is placed in an
external electrical field $\textbf{E}_{0}=E_{0}\textbf{e}_{z}$.
The values of concentration $n_{0}$ and polarization
$\textbf{P}_{0}=\kappa\textbf{E}_{0}$ for the system in an
equilibrium state are constant and uniform and corresponding values of the velocity field
$v^{\alpha}(\textbf{r},t)$ and tensor
$R^{\alpha\beta}(\textbf{r},t)$ equal to zero.

We consider small perturbations of the equilibrium state as
$$\begin{array}{ccc}n=n_{0}+\delta n,& \textbf{v}=0+\textbf{v},& \textbf{E}=\textbf{E}_{0}+\delta\textbf{E},\end{array}$$
\begin{equation}\label{di BEC equlib state BEC}\begin{array}{ccc}& & P^{\alpha}=P_{0}^{\alpha}+\delta P^{\alpha}, R^{\alpha\beta}=0+\delta R^{\alpha\beta}.\end{array}\end{equation}
Substituting these formulas in equations (\ref{di BEC
cont eq}), (\ref{di BEC bal imp eq short}), (\ref{di BEC eq
polarization}), (\ref{di BEC eq for pol current gen selfconsist
appr}) and (\ref{di BEC field eq}) \textit{and} neglecting
nonlinear terms, we obtain a system of linear homogeneous
equations in partial derivatives with constant coefficients.
Passing to the following representation for small perturbations
$\delta f$
$$\delta f =f(\omega, \textbf{k}) exp(-\imath\omega t+\imath k z) $$
yields the homogeneous system of algebraic equations, where we consider wave propagation along the external elctric field. In this case the equilibrium polarization gives maximum contribution.
The electric field strength is assumed to have a nonzero value.
Expressing all the quantities entering the system of equations in
terms of the electric field, we come to the equation
$$\Lambda\cdot \delta E_{z}=0,$$
where
$$\Lambda=\omega^{2}-\frac{\hbar^{2}k^{4}}{4m^{2}}+\frac{\Upsilon k^{2}n_{0}}{2m}+4\pi\sigma\frac{P_{0}^{2}k^{2}}{mn_{0}}$$
$$-\frac{2\pi\Upsilon k^{4}P_{0}^{2}}{m^{2}\omega^{2}-\hbar^{2}k^{4}/4+m\Upsilon k^{2}n_{0}-m\Upsilon_{2}k^{4}n_{0}}.$$ In this case, the dispersion equation is
$$\Lambda=0.$$
Solving this equation with respect to $\omega^{2}$ we obtain the
following results.

The dispersion characteristic of the CE in  the DBEC
can be expressed in the form of
$$\omega^{2}=\frac{1}{2m}\Biggl(-\frac{3}{2}\Upsilon n_{0}k^{2}+\frac{\hbar^{2}k^{4}}{2m}+\Upsilon_{2}n_{0}k^{4}+ 4\pi\sigma\frac{P_{0}^{2}k^{2}}{n_{0}}$$
\begin{equation}\label{di BEC general disp dep}\pm\sqrt{\biggl(\frac{1}{2}\Upsilon n_{0}k^{2}-\Upsilon_{2}n_{0}k^{4}+ 4\pi\sigma\frac{P_{0}^{2}k^{2}}{n_{0}}\biggr)^{2}-8\pi\Upsilon k^{4}P_{0}^{2}}\Biggr).\end{equation}
In contrast to the non-polarized BEC, where the Bogoliubov mode
exists only ~\cite{Bogoliubov N.1947,L.P.Pitaevskii RMP 99,Andreev
PRA08}, a new wave solution appears in the polarized system due to
the polarization dynamics. The Bogoliubov mode corresponds to a
solution of the equation (\ref{di BEC general disp dep}) with
minus sign before the square root. New wave solution is a wave
of polarization. In general case presented by formula (\ref{di BEC
general disp dep}) the frequency of polarization wave in the BEC
depends on $P_{0}$ and $\Upsilon$. To investigate the BEC
polarization effect on the Bogoliubov mode and the dispersion
characteristic of the new solution we analyze limit cases of the
formula (\ref{di BEC general disp dep}). If we consider wave propagation at an angle $\theta$ to the direction of the external field we find that we should put $P_{0}\cos\theta$ instead of $P_{0}$ in equation (\ref{di BEC general disp dep}).

Formula (\ref{di BEC general disp dep}) demonstrates that taking into
account of polarization dynamics in the BEC
leads to a new solution.

\begin{figure}
\includegraphics[width=8cm,angle=0]{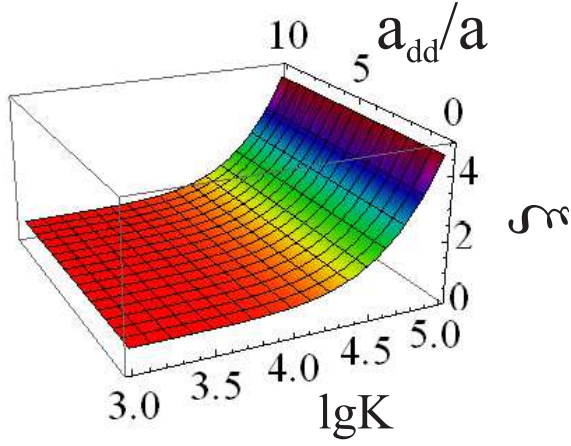}
\caption{\label{dipBEC01} (Color online) The figure presents the dependence of
reduced frequency $\xi$ on the wave vector $k$ and DSL
$a_{dd}/|a|$ for the Bogoliubov mode. We use $\lg k$, where $k$ measured
in cm$^{-1}$. Figure is obtained for the repulsive
SRI $a>0$, $\Upsilon <0$,  $|a|= 10^{-7}$ cm, $n_{0}=10^{14}$
cm$^{-3}$, $\sigma=1$, $\theta=1$. }
\end{figure}

\begin{figure}
\includegraphics[width=8cm,angle=0]{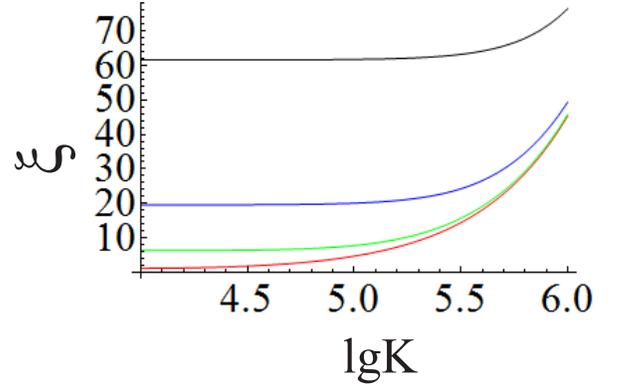}
\caption{\label{dipBEC02} (Color online) The figure presents the dependence of
reduced frequency $\xi$ on the wave vector $k$ for polarization
wave at four different values of the DSL $a_{dd}$: $a_{dd,1}=0$ (red),
$a_{dd,2}=|a|$ (green), $a_{dd,3}=10|a|$ (blue),
$a_{dd,4}=100|a|$(black). We use $\lg k$, where $k$ is measured
in cm$^{-1}$. Figure shows increasing of $\xi$ with
increasing of $a_{dd}$. Figure is obtained for repulsive SRI $a>0$,
$\Upsilon <0$, and $|a|= 10^{-7}$ cm, $n_{0}=10^{14}$ cm$^{-3}$,
$\sigma=1$, $\theta=1$. }
\end{figure}

\begin{figure}
\includegraphics[width=8cm,angle=0]{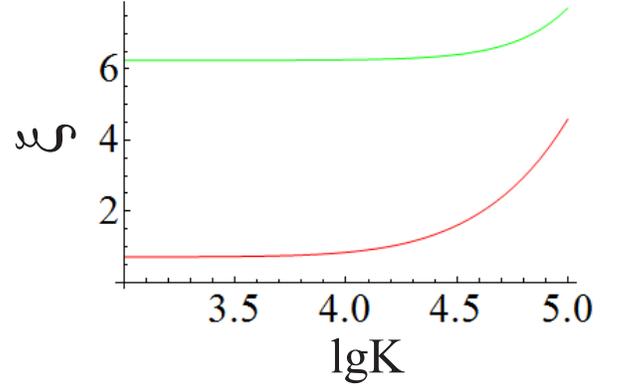}
\caption{\label{dipBEC03} (Color online) The figure shows behavior of reduced frequency
$\xi$ dependence on wave vector $k$ for both solutions presented
by formula (\ref{di BEC general disp dep special form}). We can
see the relative behavior of two solutions. Upper curve (blue)
coincides to solution with sign "+" in front of square root
(polarization mode) and lower curve (red) correspond to the one
with sign "-" (the Bogoliubov mode). The figure presents comparison of the
reduced frequencies dependence for the Bogoliubov and polarization
modes. We can see that $\xi(k)$ for polarization mode lie above the
Bogoliubov mode. We use $\lg k$, where $k$ is measured
in cm$^{-1}$. $a_{dd}=|a|$, $a>0$, $\Upsilon
<0$ (repulsion),  $|a|= 10^{-7}$ cm, $n_{0}=10^{14}$ cm$^{-3}$,
$\sigma=1$, $\theta=1$.}
\end{figure}

\begin{figure}
\includegraphics[width=8cm,angle=0]{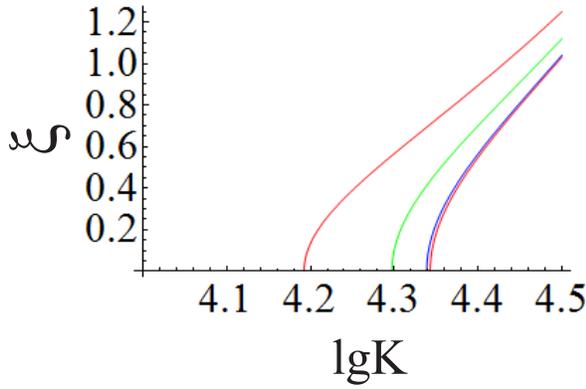}
\caption{\label{dipBEC04} (Color online) The figure presents the dependence of
reduced frequency $\xi$ on the wave vector $k$ for the Bogoliubov
mode, $a<0$, $\Upsilon >0$ (attraction), the DSL $a_{dd}=0$ (Red, the right-hand
one), $a_{dd}=0.0005|a|$ (Blue), $a_{dd}=0.005|a|$ (Green),
$a_{dd}=100|a|$ (Red, the left-hand one). We use $\lg k$, where $k$ is measured
in cm$^{-1}$. $|a|= 10^{-7}$ cm,
$n_{0}=10^{14}$ cm$^{-3}$, $\sigma=1$, $\theta=1$.}
\end{figure}

\begin{figure}
\includegraphics[width=8cm,angle=0]{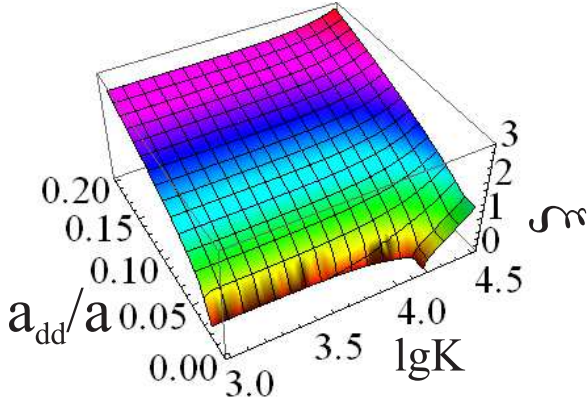}
\caption{\label{dipBEC05} (Color online) The figure presents the dependence
reduced frequency $\xi$ on the wave vector $k$ and the DSL $a_{dd}$
for polarization mode, $a<0$, $\Upsilon >0$ (attraction). We use $\lg k$, where $k$ is measured
in cm$^{-1}$. $|a|=
10^{-7}$ cm, $n_{0}=10^{14}$ cm$^{-3}$, $\sigma=1$, $\theta=1$.}
\end{figure}

\begin{figure}
\includegraphics[width=8cm,angle=0]{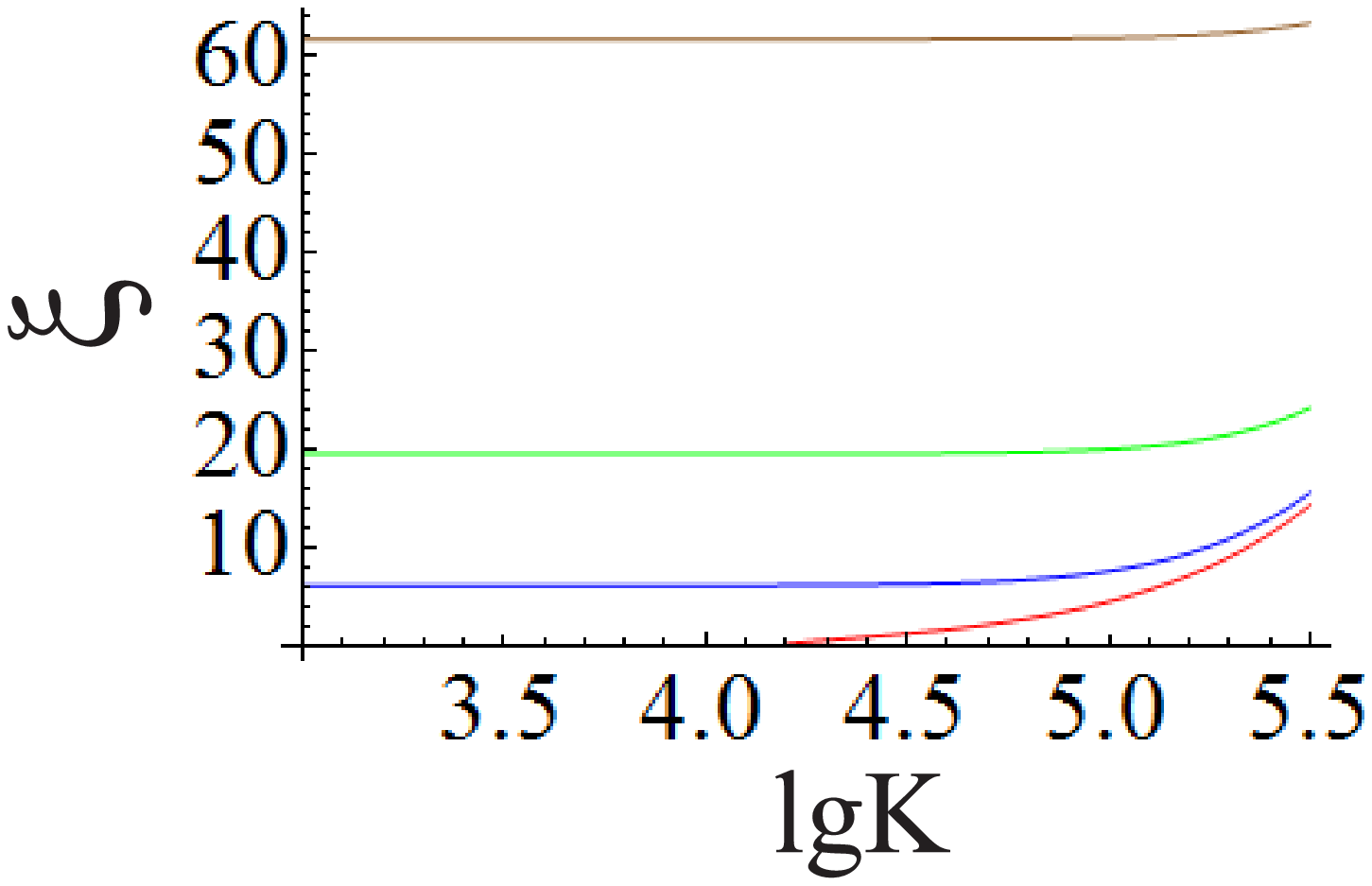}
\caption{\label{dipBEC06} (Color online) The figure presents the dependence
reduced frequency $\xi$ on the wave vector $k$ for the polarization
mode, $a<0$, $\Upsilon >0$ (attraction), DSL $a_{dd}=0$ (Red),
$a_{dd}=|a|$ (Blue), $a_{dd}=10|a|$ (Green), $a_{dd}=100|a|$ (Brown).
We use $\lg k$, where $k$ is measured
in cm$^{-1}$.
$|a|= 10^{-7}$ cm, $n_{0}=10^{14}$ cm$^{-3}$, $\sigma=1$,
$\theta=1$.}
\end{figure}

Let us start with the case when contributions of different term in
formula (\ref{di BEC general disp dep}) are comparable. In this
case we study dispersion numerically. For this purpose we
represent formula (\ref{di BEC general disp dep}) in the following
form
$$\xi^{2}=\biggl(\frac{3}{4}\frac{a}{|a|}+\frac{k^{2}}{16\pi n_{0}|a|}-\frac{\theta a^{3}k^{2}}{16|a|}+\frac{6\pi\sigma a_{dd}}{|a|}$$
\begin{equation}\label{di BEC general disp dep special form}\pm\sqrt{\biggl(-\frac{1}{4}\frac{a}{|a|}+\frac{\theta a^{3}k^{2}}{16|a|}+\frac{6\pi\sigma a_{dd}}{|a|}\biggr)^{2}+ \frac{6\pi a_{dd}}{a}}\biggr),\end{equation}
where
$$\xi^{2}\equiv\frac{\omega^{2}}{|\Upsilon| n_{0}k^{2}/m},$$
and we use the length $a_{dd}$ measures the strength of dipolar
interaction, and defined by formula
$$a_{dd}=\frac{mP_{0}^{2}}{12\pi n_{0}^{2}\hbar^{2}}.$$

Above we have discussed that we do not consider the scattering process,
but analogy with the results of other papers we
introduced an effective parameter $a_{dd}$ describing contribution of the DDI, although $a_{dd}$ does not connect with
scattering.

In the following we describe a numerical analysis of formula
(\ref{di BEC general disp dep special form}). Reduced frequency
$\xi$ for two waves presented by formula (\ref{di BEC general disp
dep special form}) is exhibited on Fig.s \ref{dipBEC01} and
\ref{dipBEC02}. They are obtained for the repulsive SRI ($a>0$,
$\Upsilon <0$) at $a=10^{-7}$cm and $n_{0}=10^{14}$cm$^{-3}$.
These Fig.s show the reduced frequency $\xi(k)$ for different
values of equilibrium polarization and, consequently, for
different DSL $a_{dd}$.

Fig. \ref{dipBEC01} presents solution with sign "-" in front of the
square root in formula (\ref{di BEC general disp dep}), this
solution corresponds to the Bogoliubov mode. For all $a_{dd}/
|a|$ the $\xi(k)$ does not depend on $a_{dd}/|a|$.

Fig. \ref{dipBEC02} shows strong dependence of the reduced frequency of the
polarization mode ("+" in front of the square root in formula (\ref{di BEC
general disp dep})) on equilibrium polarization. On
Fig.\ref{dipBEC02} as on other Fig.s this dependence is presented via
dependence on the DSL divided on module of the SL of SRI $a_{dd}/|a|$.

The form of $\xi(k)$ for two solutions has similar form. We
present both solutions on Fig. \ref{dipBEC03} at fixed
$a_{dd}/|a|=1$ to compare. From Fig. \ref{dipBEC03} we see that the
reduced frequency of polarization mode is larger than one for
Bogoliubov mode.

From the Fig.\ref{dipBEC02} we obtain that
polarization lead to growing of the reduced frequency $\xi$ of the polarization mode. The reduced frequency $\xi$ increases in 6 times at
small wave vectors of order $10^{4}$cm$^{-1}$ and for the
$a_{dd}=|a|$. For
the $a_{dd}=100|a|$ increases in 60 times. With increasing
of wave vector the influence of polarization becames smaller. For
example at $k=10^{6}$cm$^{-1}$ and $a_{dd}=100|a|$ the reduced
frequency $\xi$ increase just in 1.6 times.

For the case of an attractive SRI in the absence of the polarization
there is instability of the Bogoliubov mode at small wave vectors.
At the presence of small polarization the Bogoliubov mode
remains unstable, but the area of stability become some wider as it is
shown on Fig. \ref{dipBEC04}. There is fast widening of the region of
stability at increasing of $a_{dd}/|a|$ Fig. \ref{dipBEC05}.
Widening starts at approximately $a_{dd}=0.02|a|$.

For the case of the attractive SRI in the absence of polarization
there is instability of polarization mode at small wave vectors
(see Fig. \ref{dipBEC06}). At $a_{dd}=0$ we see instability of
waves at wave vector $k$ below $1.6\cdot 10^{4}$cm$^{-1}$.
The EDM leads to increasing of $\xi$ and fast growing of
the region of stability. The wave is stable up to $k=10^{3}$cm$^{-1}$
at $a_{dd}/|a|=1$. Further increasing of $a_{dd}/|a|$ leads to
increasing of $\xi$ up to 64. $\xi$ become equal to 64 at
$a_{dd}/|a|\simeq 100$ and does not change at further increasing
of $a_{dd}/|a|$.

Comparing Fig.s \ref{dipBEC04} and \ref{dipBEC06} we can see that
reduced frequency $\xi$ of polarization mode much larger than the
Bogoliubov mode, as it was for the repulsive SRI.

\begin{figure}
\includegraphics[width=8cm,angle=0]{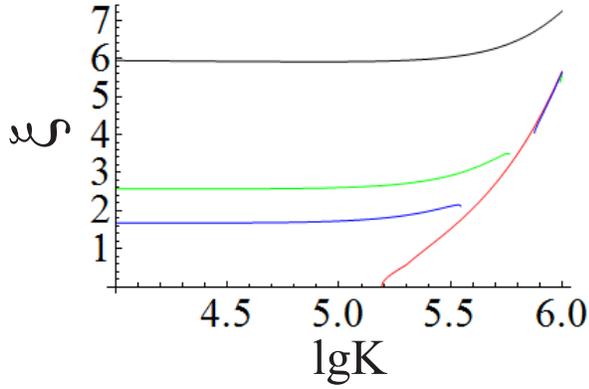}
\caption{\label{dipBEC01TOIR} (Color online) The figure presents the dependence of the
reduced frequency $\xi$ on the wave vector $k$ for the polarization
mode. On this Fig. the region of instabilities is also shown. It
is a region, where we see a gap in curves. Figure obtained for the attractive
SRI ($a<0$, $\Upsilon >0$),  $a_{dd}=0$ (Red), $a_{dd}=0.1
|a|$ (Blue), $a_{dd}=0.2 |a|$ (Green), $a_{dd}= |a|$ (Black). We use $\lg k$, where $k$ is measured
in cm$^{-1}$.
$|a|$= 10$^{-5}$ cm, $n_{0}=10^{14}$ cm$^{-3}$, $\sigma=1$, $\theta=1$. %, red --, blue ++
}
\end{figure}

\begin{figure}
\includegraphics[width=8cm,angle=0]{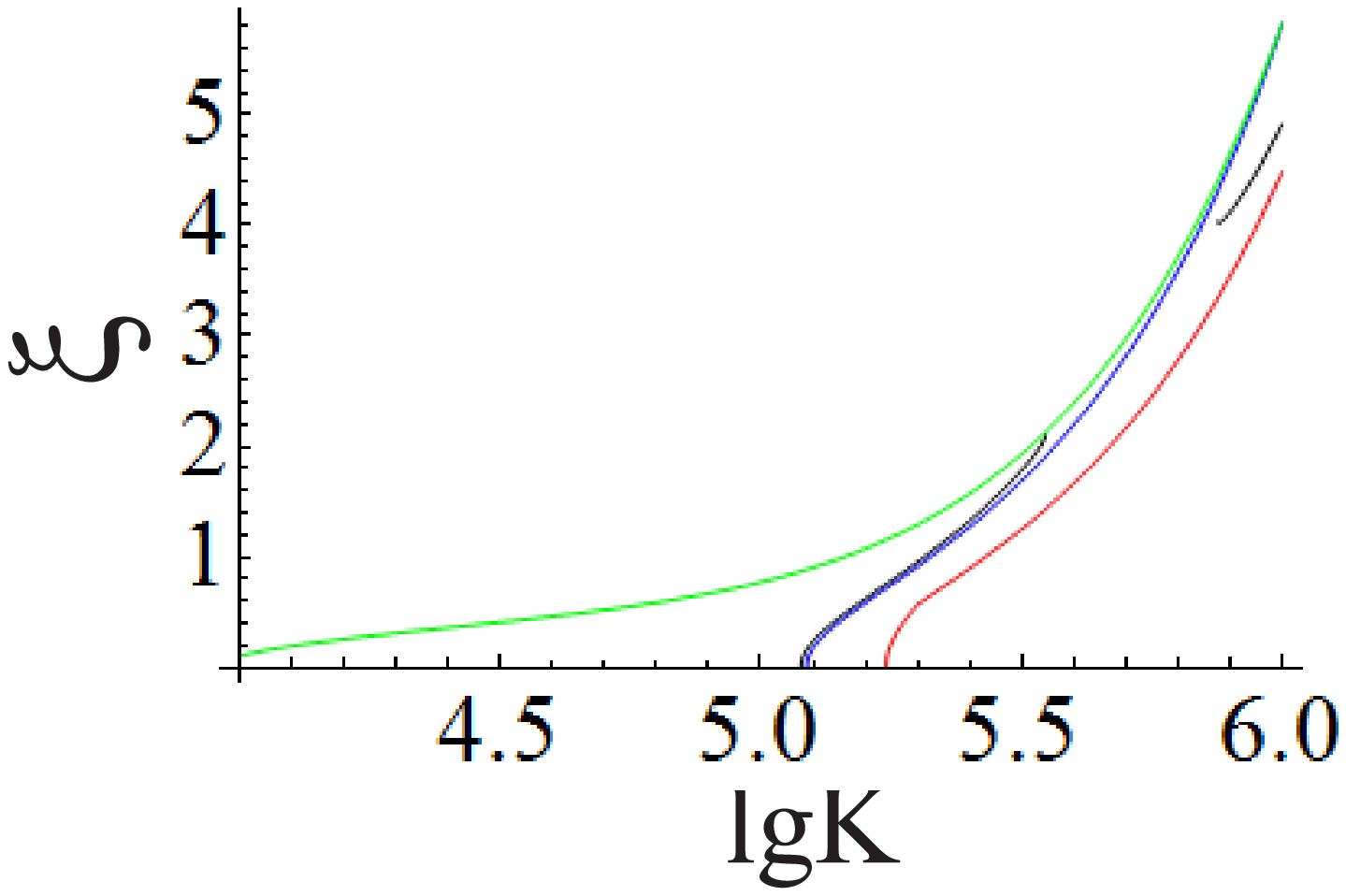}
\caption{\label{dipBEC02TOIR} (Color online) The figure presents the dependence of the
reduced frequency $\xi$ on the wave vector $k$ for the Bogoliubov
mode, $a<0$, $\Upsilon >0$ (attraction), the DSL $a_{dd}=0$ (Red),
$a_{dd}=
0.1|a|$ (Black), $a_{dd}= 0.2|a|$ (Blue), $a_{dd}= 2|a|$ (Green). We use $\lg k$, where $k$ is measured
in cm$^{-1}$.
$|a|$= 10$^{-5}$ cm, $n_{0}=10^{14}$ cm$^{-3}$, $\sigma=1$, $\theta=1$.}%--
\end{figure}

\begin{figure}
\includegraphics[width=8cm,angle=0]{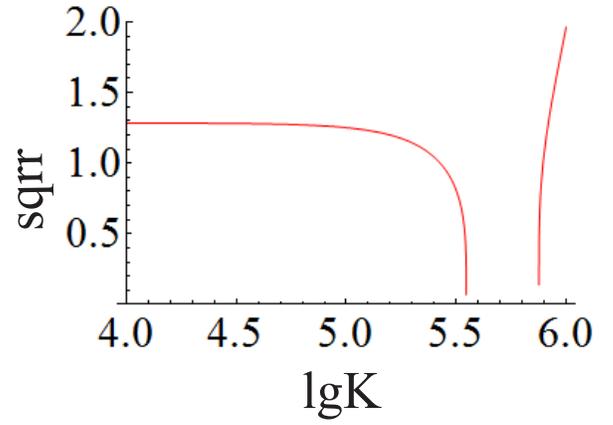}
\caption{\label{dipBEC03TOIR} (Color online) The figure presents the behavior of
the square root in formula (\ref{di BEC disp dependence with the
beam simpl}) (we conditionally designated this quantity as "sqrr") at different values of the wave vector $k$ corresponding to
the Fig. \ref{dipBEC04} at $a_{dd}=0.1|a|$, (square does not
depend on concentration). $|a|$= 10$^{-5}$ cm, $\sigma=1$,
$\theta=1$.  We use $\lg k$, where $k$ is measured
in cm$^{-1}$.
}%++
\end{figure}

With increasing of strength of the SRI the role of $\Upsilon_{2}$
become more important. It's influence reveals at $|a|=10^{-5}$cm
for the attractive SRI. In this case, there is  instability region at
relatively large wave vectors. For described case the reduced
frequency  $\xi(k)$ is presented on Fig.s \ref{dipBEC01TOIR} and
\ref{dipBEC02TOIR} at various $a_{dd}/|a|$. The region of
instability is shown there. This instability appears because the
expression under the square root in formula (\ref{di BEC general
disp dep}) became negative as it shown on Fig. \ref{dipBEC03TOIR}.


Below, we analytically  consider limit cases of general dispersion
solution (\ref{di BEC general disp dep}).

We now proceed to derive a dispersion dependence when the effect of polarization is small in
comparison with the contribution of the short-range effects, i.e. we consider a limit when the
terms proportional to $\Upsilon$ and $\Upsilon_{2}$ are comparably large. Consequently,
formula (\ref{di BEC general disp dep}) takes the form
$$\omega^{2}_{B}=\frac{\hbar^{2}k^{4}}{4m^{2}}-\frac{\Upsilon n_{0}k^{2}}{m}+\frac{\Upsilon_{2}n_{0}k^{4}}{m}$$
\begin{equation}\label{di BEC 01a}-\frac{4\pi P_{0}^{2}k^{2}}{mn_{0}}\frac{\Upsilon(\sigma-1)-2\sigma\Upsilon_{2}k^{2}}{\Upsilon-2\Upsilon_{2}k^{2}},\end{equation}
and
$$\omega^{2}_{P}=\frac{\hbar^{2}k^{4}}{4m^{2}}-\frac{1}{2}\frac{\Upsilon
n_{0}k^{2}}{m}$$
\begin{equation}\label{di BEC 01b}+\frac{4\pi P_{0}^{2}k^{2}}{mn_{0}}\frac{\Upsilon(\sigma-1)-2\sigma\Upsilon_{2}k^{2}}{\Upsilon-2\Upsilon_{2}k^{2}},\end{equation}
where we use indexes "B" for the Bogoliubov mode and "P" for new
polarization mode. At derivation of formulas (\ref{di BEC 01a}) and
(\ref{di BEC 01b}) we expanded a sub-radical expression in
(\ref{di BEC general disp dep}) and took only first two terms of
the expansion to estimate the influence of polarization on the
wave dispersion.

As it follows from equations (\ref{di BEC eq polarization}) and
(\ref{di BEC eq for pol current gen selfconsist appr}), changes of
polarization can occur due to the dipole-dipole interactions, the SRI and the quantum Bohm potential (terms proportional
to $\hbar^{2}$). This is the reason for existence of other ways of
polarization changing when the contribution of equilibrium
polarization $P_{0}$ is negligible. In described case, changes in the polarization do not affect
the concentration evolution in
linear approximation.

Formula (\ref{di BEC general disp dep}) is valid even in
absence of the external electrical field when equilibrium
polarization equals to zero $P_{0}=0$. In this case, equations (\ref{di BEC general
disp dep}) appears as
\begin{equation}\label{di BEC disp in absence ext field a}\omega^{2}_{B}=\frac{1}{m}\Biggl(\frac{\hbar^{2}k^{4}}{4m}-\Upsilon n_{0}k^{2}+\Upsilon_{2}n_{0}k^{4}\Biggr),\end{equation}
and
\begin{equation}\label{di BEC disp in absence ext field b}\omega^{2}_{P}=\frac{1}{2m}\Biggl(\frac{\hbar^{2}k^{4}}{2m}-\Upsilon n_{0}k^{2}\Biggr).\end{equation}

The waves of polarization can exist at the absence of
external electric field $E_{0}=0$. In this case the equilibrium
state is not polarized and dipole direction of particles is
distributed accidentally.

If the contribution of equilibrium polarization in the BEC
dispersion (\ref{di BEC general disp dep}) is comparable to the
contribution of the SRI in the third order by the
interaction radius  $\Upsilon_{2}$ (\ref{di BEC
general disp dep}) transforms into
\begin{equation}\label{di BEC 02a}\omega^{2}_{B}=\frac{\hbar^{2}k^{4}}{4m^{2}}-\frac{\Upsilon n_{0}k^{2}}{m}+\frac{\Upsilon_{2}n_{0}k^{4}}{m}+4\pi\frac{P_{0}^{2}k^{2}}{mn_{0}},\end{equation}
and
\begin{equation}\label{di BEC 02b}\omega^{2}_{P}=\frac{\hbar^{2}k^{4}}{4m^{2}}-\frac{1}{2}\frac{\Upsilon n_{0}k^{2}}{m}+4\pi\frac{P_{0}^{2}k^{2}}{mn_{0}}(\sigma-1).\end{equation}

Using Feshbach resonance ~\cite{Chin RMP 10,Bloch RMP 08} we can
tune the SRI potential that $\Upsilon=0$ while
$\Upsilon_{2}\neq 0$. In
this situation formula (\ref{di BEC general disp dep}) turns into
\begin{equation}\label{di BEC 03a}\omega^{2}_{B}=\frac{\hbar^{2}k^{4}}{4m^{2}}+\frac{\Upsilon_{2}n_{0}k^{4}}{m},\end{equation}
and
\begin{equation}\label{di BEC 03b}\omega^{2}_{P}=\frac{\hbar^{2}k^{4}}{4m^{2}}+4\pi\sigma\frac{P_{0}^{2}k^{2}}{mn_{0}}.\end{equation}

In this paper we primarily focus on the influence of the BEC
polarization on its dispersion characteristics. So, let us consider
the case when the contribution into the dispersion of the
SRI at the first order of the interaction
radius, i.e. terms proportional to $\Upsilon$, is comparable to the
contribution of polarization, and their total effect is much
larger than the contribution of terms proportional to
$\Upsilon_{2}$. We find that formula (\ref{di BEC general disp dep}) turns into
\begin{equation}\label{di BEC 05a}\omega^{2}_{B}=\frac{\hbar^{2}k^{4}}{4m^{2}}-\frac{\Upsilon n_{0}k^{2}}{m}\biggl(\frac{(\sigma-1)}{\sigma}\biggr),\end{equation}
and
$$\omega^{2}_{P}=\frac{\hbar^{2}k^{4}}{4m^{2}}+4\pi\sigma\frac{P_{0}^{2}k^{2}}{mn_{0}}$$
\begin{equation}\label{di BEC 05b}-\frac{1}{2}\frac{\Upsilon n_{0}k^{2}}{m}\biggl(\frac{(\sigma+4)}{2\sigma}\biggr).\end{equation}

\subsection{polarization evolution at constant concentration}

The DBEC is the rare gas and we can not use condition
incompressibility, but we consider the case there concentration of
particles is not change in time. We are interested in dynamics of wave
of polarization only. For this aim we need to use equations
(\ref{di BEC eq polarization}) and (\ref{di BEC eq for pol current
gen selfconsist appr}) at additional condition
$n(\textbf{r},t)=n_{0}=const$. This is means the polarization
changes due to evolution of dipoles direction. Equation of
polarization evolution (\ref{di BEC eq polarization}) has no
change in this approximation. Nonlinear terms in equation (\ref{di BEC eq for pol current gen
selfconsist appr}) disappear and the term
describing the SRI simplifies, so we have
$$\partial_{t}R^{\alpha\beta}(\textbf{r},t)+\partial^{\gamma}\biggl(R^{\alpha\beta}(\textbf{r},t)v^{\gamma}(\textbf{r},t)$$
$$+R^{\alpha\gamma}(\textbf{r},t)v^{\beta}(\textbf{r},t)-P^{\alpha}(\textbf{r},t)v^{\beta}(\textbf{r},t)v^{\gamma}(\textbf{r},t)\biggr)$$
$$-\frac{\hbar^{2}}{4m^{2}}\partial_{\beta}\triangle P^{\alpha}(\textbf{r},t)=\frac{n_{0}}{m}\Upsilon\partial^{\beta}\biggl(P^{\alpha}(\textbf{r},t)\biggr)$$
\begin{equation}\label{di BEC eq for pol current gen selfconsist
appr simpl
}+\frac{\sigma}{mn_{0}}P^{\alpha}(\textbf{r},t)P^{\gamma}(\textbf{r},t)\partial^{\beta}E^{\gamma}(\textbf{r},t).\end{equation}
in the set with equations of field (\ref{di BEC field eq}) and (\ref{di BEC field eq rot}).

Only
one wave solution exists in this approximation, it is the wave of polarization. Its dispersion
is
\begin{equation}\label{di BEC disp of pol only}\omega^{2}=\frac{\hbar^{2}k^{4}}{4m^{2}}+\biggl(\frac{4\pi\sigma P_{0}^{2}}{mn_{0}}-\frac{n_{0}\Upsilon }{m}\biggr)k^{2}.\end{equation}
Formula (\ref{di BEC disp of pol only}) reduces to
$$\omega^{2}=\frac{4\pi\sigma P_{0}^{2}}{mn_{0}}k^{2}$$
than polarization dominates. For the particles with the EDM equal to 1 Debye at concentration
10$^{12}$ cm$^{-3}$ and wave vector $k\simeq 10^{7}$ cm$^{-1}$ the
frequency $\omega$ reach value of order $10^{6}-10^{7}$ s$^{-1}$.

Dispersion of the 1D and 2D wave of polarization is described in Ref. ~\cite{Andreev PRB 11}.

\section{Generation of waves in polarized BEC}

In this section we consider the process of wave generation in the BEC
by means of the beam of neutral polarized particles. The
interaction between the beam and the BEC has the dipole-dipole origin.

 To get the dispersion solution  we use the system of the QHD
equations for each sort of particles (\ref{di BEC cont eq}),
(\ref{di BEC bal imp eq short}), (\ref{di BEC eq polarization}),
(\ref{di BEC eq for pol current gen selfconsist appr}) and the
equations of field (\ref{di BEC field eq}) and (\ref{di BEC field eq rot}). The equilibrium state
of system is characterized by following values of the BEC
parameters:
$$\begin{array}{ccc}n_{B}=n_{B0}+\delta n_{B},& v^{\alpha}_{B}=0+v^{\alpha}_{B},& \end{array}
$$
\begin{equation}\label{di BEC equlib state BEC}\begin{array}{ccc}& P^{\alpha}_{B}=P_{B0}^{\alpha}+\delta P^{\alpha}_{B},& R^{\alpha\beta}_{B}=0+\delta R^{\alpha\beta}_{B},\end{array}
\end{equation}
and values of the beam parameters:
$$\begin{array}{ccc}n_{b}=n_{0b}+\delta n_{b},& v^{\alpha}_{b}=U\delta^{z\alpha}+\delta v^{\alpha}_{b},& \end{array}
$$
\begin{equation}\label{di BEC equlib state beam}\begin{array}{ccc} & P_{b}^{\alpha}=P_{0b}^{\alpha}+\delta P^{\alpha}_{b},& R^{\alpha\beta}_{b}=R^{\alpha\beta}_{0b}+\delta R^{\alpha\beta}_{b}.\end{array}
\end{equation}
The polarization $P_{0}^{\alpha}$ is proportional to the external
electric field $E_{0}^{\alpha}$. We consider the case then
$\textbf{E}_{0}=[E_{0}sin\varphi, 0, E_{0}cos\varphi]$. In this
case the tensor $R^{\alpha\beta}_{0b}$ has only two
nonzero elements: $R^{zx}_{0b}=R_{0b}sin\varphi$ and
$R^{zz}_{0b}=R_{0b}cos\varphi$.
For the process, under consideration the dispersion relation is:
$$1+4\pi k^{2}\Biggl(\frac{P_{0}^{2}}{\omega^{2}-\frac{\hbar^{2}k^{4}}{4m^{2}}+\frac{\Upsilon n_{0}k^{2}}{2m}}\times$$
$$\times\biggl(\frac{\Upsilon k^{2}/(2m^{2})}{\omega^{2}-\frac{\hbar^{2}k^{4}}{4m^{2}}+\frac{\Upsilon n_{0}k^{2}}{m}-\frac{\Upsilon_{2}n_{0}k^{4}}{m}}-\frac{\sigma}{mn_{0}}\biggr) $$
$$+\frac{1}{(\omega-k_{z}U)^{2}-\frac{\hbar^{2}k^{4}}{4m^{2}_{b}}}\times$$
\begin{equation}\label{di BEC disp dependence with the beam} \times\biggl(\frac{2(\omega-k_{z}U)P_{0b}k_{z}(P_{0b}U-R_{0b})}{m_{b}n_{0b}\biggl((\omega-k_{z}U)^{2}-\frac{\hbar^{2}k^{4}}{4m^{2}_{b}}\biggr)}-\frac{\sigma_{b}P_{0b}^{2}}{m_{b}n_{0b}}\biggr)\Biggr)=0.\end{equation}
Using relation $P_{0b}U-R_{0b}=0$, we can simplify the equation
(\ref{di BEC disp dependence with the beam}) and obtain

$$1+\frac{\omega_{D}^{2}}{\omega^{2}-\omega_{1}^{2}}\Biggl(\frac{\Upsilon n_{0}k^{2}/(2m)}{\omega^{2}-\omega_{2}^{2}}-\sigma\Biggr)$$
\begin{equation}\label{di BEC disp dependence with the beam simpl}-\frac{\sigma_{b}\omega_{Db}^{2}}{(\omega-k_{z}U)^{2}-\frac{\hbar^{2}k^{4}}{4m^{2}_{b}}}=0.\end{equation}
In this formula the following designations are used
\begin{equation}\label{di BEC omega 1}\omega_{1}^{2}=\frac{\hbar^{2}k^{4}}{4m^{2}}-\frac{\Upsilon n_{0}k^{2}}{2m},\end{equation}
\begin{equation}\label{di BEC omega 2}\omega_{2}^{2}=\frac{\hbar^{2}k^{4}}{4m^{2}}-\frac{\Upsilon n_{0}k^{2}}{m}+\frac{\Upsilon_{2}n_{0}k^{4}}{m}\end{equation}
and
\begin{equation}\label{di BEC omega D}\omega_{Di}^{2}=\frac{4\pi P_{0i}^{2}k^{2}}{m_{i}n_{0i}},\end{equation}
where $i$ is the index of sorts of particles, the BEC ($i=B$) or the beam ($i=b$).

The equation (\ref{di BEC disp dependence with the beam simpl})
has two beam related solutions, in the absence of BEC medium:
\begin{equation}\label{di BEC beam mode 1}\omega=k_{z}U\pm\sqrt{\frac{\hbar^{2}k^{4}}{4m^{2}_{b}}+\sigma_{b}\omega_{Db}^{2}}.\end{equation}
We will consider the possibilities of instability for the case
of low-density beam $\omega_{Db}\sim
n_{0b}\rightarrow 0$. In this case we can neglect the last term under the
square root in (\ref{di BEC beam mode 1}).  The resonance
interaction of beam with the BEC realizing at
\begin{equation}k_{z}U\pm\frac{\hbar k^{2}}{2m_{b}}=\omega(k),\end{equation}
and could lead to instabilities. The quantity $\omega(k)$ is the
dispersion of BEC modes (\ref{di BEC general disp dep}).  The
frequency in this case can be presented in the form
\begin{equation}\label{di BEC influence of beam mode}\omega=k_{z}U\pm\frac{\hbar k^{2}}{2m_{b}}+\delta\omega.\end{equation}
Let us to consider two limit cases.

\subsection{small frequency shift limit}

Under condition
\begin{equation}\label{di
BEC}\delta\omega\ll\hbar k^{2}/m\end{equation}
the frequency shift
emerges as
\begin{equation}\label{di BEC increment} \delta\omega^{2}=\pm\frac{2\sigma_{b}m_{b}m^{2}\omega_{Db}^{2}(\omega^{2}-\omega_{1}^{2})^{2}(\omega^{2}-\omega_{2}^{2})^{2}}{\omega\omega_{D}^{2}\hbar\Upsilon^{2}n_{0}^{2}k^{6} W}, \end{equation}
where
\begin{equation}\label{di BEC W general}W=2\omega^{2}-\omega_{1}^{2}-\omega_{2}^{2}-\frac{2m\sigma}{\Upsilon n_{0}k^{2}}(\omega^{2}-\omega_{2}^{2})^{2}.\end{equation}
and the frequency $\omega$ determined with formula (\ref{di BEC
general disp dep}). The instabilities take place when
$\delta\omega^{2}<0$. The sign of $\delta\omega^{2}$ depends on
the sign of $W$.

For the case resonance interaction of beam with the waves in BEC
there are instabilities, for the first beam mode in (\ref{di BEC
beam mode 1}) at $W<0$ and for the second beam mode in formula
(\ref{di BEC beam mode 1}) at $W>0$. For the polarization mode $W$
is positive. It means that interaction of polarization mode
with the second beam related mode results in the instability. For
the Bogoliubov mode the sign of $W$ depends on $\sigma$.

We can consider the following cases:

(i) the contribution of equilibrium polarization to $\omega(k)$ is
dominant; then $W>0$;

(ii) main contribution in $\omega(k)$ gives term proportional to $\Upsilon$,
\textit{and}, equilibrium polarization and SRI in the TOIR approximation are
comparable to each other. In this case, there is a value of $\sigma$, designated as $\sigma_{0}$, which is
$$\sigma_{0}=1+\frac{\Upsilon_{2}n_{0}k^{4}/m}{\Upsilon
n_{0}k^{2}/(2m)+2\Upsilon_{2}n_{0}k^{4}/m-4\omega_{D}^{2}}.$$
The
sign of $W$ varies at $\sigma=\sigma_{0}$. The dependence of sign of $W$
on $\sigma$ is presented in the Table 1.

\begin{table}[b]\caption{\label{tab:table1}%
In this table the sign of the $W$ are presented for the
Bogoliubov mode when the SRI is prevailed over dipole-dipole interaction.}
%\begin{ruledtabular}
\begin{tabular}{lcr}
 \textrm{ }&
 \textrm{\large $\frac{\Upsilon>0}{\textmd{attraction}}$}&
\multicolumn{1}{c}{\textrm{\large $\frac{\Upsilon<0}{\textmd{repulsion}}$}}\\
%\colrule
 & & \\
  &  &  \\
\large
$\sigma>\sigma_{0}$ & \large + & \large -\\
 &  &  \\
\large $\sigma<\sigma_{0}$ & \large - & \large +\\
\end{tabular}
%\end{ruledtabular}
\end{table}

\subsection{large frequency shift limit}

In opposite limit
\begin{equation}\label{di
BEC large limit}\delta\omega\gg\hbar k^{2}/m\end{equation}
we have
\begin{equation}\label{di BEC W rep in pol wave Q}\delta\omega  =\xi\sqrt[3]{\frac{\sigma_{b}m^{2}\omega_{Db}^{2}(\omega^{2}-\omega_{1}^{2})^{2}(\omega^{2}-\omega_{2}^{2})^{2}}{\omega\omega_{D}^{2}\Upsilon^{2}n_{0}^{2}k^{4}\mid W\mid}},\end{equation}
where $\xi$ equal to $\xi_{1}=\sqrt[3]{1}$ for $W>0$ or
$\xi_{-1}=\sqrt[3]{-1}$ for $W<0$. Explicit form of quantities
$\xi_{1}$ and $\xi_{-1}$ are
$$\xi_{-1}=[-1, \frac{-1+\imath\sqrt{3}}{2}, \frac{-1-\imath\sqrt{3}}{2}],$$
and
$$\xi_{1}=[1, \frac{1+\imath\sqrt{3}}{2}, \frac{1-\imath\sqrt{3}}{2}].$$

The consideration concerning the sign of $W$, reflected in the
table 1, are valid also for the limit condition (\ref{di BEC large
limit}).

From the formulas (\ref{di BEC increment}) and (\ref{di BEC W rep
in pol wave Q}) we can see that neutral particles beam
leads to instabilities for both the Bogoliubov and polarization
waves.

\section{Conclusion}

We developed the self-consistent method for
description of the DBEC dynamics. This method accounts
effect of polarization on changes of the particle concentration and
velocity field, which are determined in general by the continuity
equation and Euler equation. We derived the evolution equations
of the polarization and the polarization current. The derived equations
contain information about the influence of the
interactions on the polarization evolution. We studied the effect
of polarization on the BEC dynamics and influence of the SRI on
the polarization evolution. An expression of the SRI contribution in
the equation of polarization current evolution via concentration,
polarization and the SRI potential $\Upsilon=\Upsilon(U_{ij})$ was
derived. With the assumption that the state of polarized Bose
particles in the form of BEC can be described with some
single-particle wave function. Changes in polarization due to SRI
are shown to be determined at the first order of the interaction
radius by the same interaction constant that occurs in Euler's
equation and Gross-Pitaevskii equation.

The derivation of the GP equation for polarized particles from the QHD
equations was obtained. The conditions of validity of the GP
equation was presented. Comparison of evolution equation of the
electrically polarized BEC and the magnetized BEC was discussed and
differences were described.

Physical meaning of the self-consistent approximation was
described. Suitability of the self-consistent approximation for
electrically polarized BEC was shown. Distinctions of developed self-consistent approximation for dipole-dipole
interaction from the scattering process are discussed.

Correct form of the dipole-dipole interaction Hamiltonian is
discussed. The arguments for choosing of correct form of
Hamiltonian for electric and magnetic dipoles are presented.

The dispersion of the CE in the polarized BEC was analyzed. We shown that polarization evolution in the BEC causes a novel
type of waves. The effect of polarization on the dispersion
of the Bogoliubov mode and the dispersion of a new wave mode
were studied.

We shown the possibility of wave generation in the polarized BEC by
means of the monoenergetic beam of neutral polarized particles.

\section{Appendix A}

\subsection{Correct form of Hamiltonian for dipole-dipole interaction}

At derivation of the QHD equations we need to write explicit
form of the Hamiltonian of dipole-dipole interaction. In some works
the dynamics of the magnetic dipole moment and of the EDM
~\cite{Lahaye RPP 09,Yi PRA 02} are analyzed in similar
ways. Usual expression of the Hamiltonian for the dipole-dipole
interaction for the electric and the magnetic dipoles equals
\begin{equation}\label{di BEC pot of dd int} H_{dd}=\frac{\delta^{\alpha\beta}-3r^{\alpha}r^{\beta}/r^{2}}{r^{3}}d_{1}^{\alpha}d_{2}^{\beta}.\end{equation}
Generally speaking the Hamiltonian for the spin-spin and the electric dipole-dipole interaction are different, and they also differ from (\ref{di BEC pot of dd int}). All Hamiltonians contain a term proportional to the Dirac delta function. These terms have different numerical coefficient for two kind of BECs. In this paper for interaction of electric dipoles we used following Hamiltonian
$$H_{dd}=-\partial^{\alpha}\partial^{\beta}\frac{1}{r}\cdot d_{1}^{\alpha}d_{2}^{\beta},$$
which follows from potential of the electric field caused by electric dipole \cite{Landau 2}.
There is well-known identity
\begin{equation}\label{di BEC togdestvo}-\partial^{\alpha}\partial^{\beta}\frac{1}{r}= \frac{\delta^{\alpha\beta}-3r^{\alpha}r^{\beta}/r^{2}}{r^{3}}+\frac{4\pi}{3}\delta^{\alpha\beta}\delta(\textbf{r}),\end{equation}
so we can see the difference between usually using Hamiltonian and one's used in this paper. The corrections of our selection followed from the fact that the equations obtained in the paper coincide to the Maxwell equations. Derivation of the QHD equations in the self-consistent approximation leads to field equation (\ref{di BEC field eq}) and (\ref{di BEC field eq rot}), we rewrite them here
\begin{equation}\label{di BEC field good}\nabla\textbf{E}(\textbf{r},t)=-4\pi \nabla\textbf{P}(\textbf{r},t),\end{equation}
and
\begin{equation}\label{di BEC field good rot}\nabla\times\textbf{E}(\textbf{r},t)=0,\end{equation}
but if we used Hamiltonian (\ref{di BEC pot of dd int}) we  would obtain
\begin{equation}\label{di BEC field wrong}\nabla\textbf{E}(\textbf{r},t)=\frac{8\pi}{3} \nabla\textbf{P}(\textbf{r},t),\end{equation}
instead of (\ref{di BEC field good}). Using formula (\ref{di BEC pot of dd int}) leads to non-fulfilment of equation (\ref{di BEC field good rot}).

It has been shown by Breit ~\cite{spin-spin interaction}
that the Hamiltonian for the spin-spin interaction contains a
term proportional to the Dirac $\delta$-function
$\delta(r_{1}-r_{2})d_{1}^{\alpha}d_{2}^{\beta}$ along with (\ref{di BEC pot of dd int}). The coefficient
at the $\delta$ function was refined later ~\cite{MaksimovTMP
2001}, so it was shown that the Hamiltonian is in according with the Maxwell equations (\ref{di BEC field good}), (\ref{di BEC field good rot}). The conclusive expression for the spin-spin interaction Hamiltonian appears as
\begin{equation}\label{di BEC pot of mm int} H_{\mu\mu}=-\Biggl(4\pi\delta^{\alpha\beta}\delta(\textbf{r}_{12})+\nabla^{\alpha}_{1}\nabla^{\beta}_{1}(1/r_{12})\Biggr)\mu^{\alpha}_{1}\mu^{\beta}_{2},\end{equation}
or
\begin{equation}\label{di BEC pot of mm int2}H_{\mu\mu}=\Biggl(\frac{\delta^{\alpha\beta}-3r^{\alpha}r^{\beta}/r^{2}}{r^{3}}-\frac{8\pi}{3}\delta^{\alpha\beta}\delta(\textbf{r})\Biggr)\mu^{\alpha}_{1}\mu^{\beta}_{2}.\end{equation}

Thus, the differences in the dipole-dipole interactions of the
electric dipoles, and the magnetic dipoles, have to be taken into account
at development of theoretical field apparatus.

\subsection{Method of equations derivation}

The Schr\"{o}dinger equation defines wave function in the
3N-dimensional configuration space while physical processes in systems
that involve large number of particles occur in the three-dimensional
physical space ~\cite{Goldstein}. This is why a problem evolves
obtaining of the quantum-mechanical description of a system of
particles in terms of material fields, e.g. the particle concentration, the momentum
density, the energy density and other fields of various tensor
dimension that are defined in the three-dimension space.

The first step in the many-particle QHD equation derivation is the definition of particles concentration. We define the quantum particles concentration as the quantum average of the classic microscopic concentration in the coordinate representation on many-particle wave function $\psi(R,t)$. Thus, the particles concentration explicit form to be
$$n(\textbf{r},t)=\int
 dR\sum_{i}\delta(\textbf{r}-\textbf{r}_{i})\psi^{*}(R,t)\psi(R,t),$$
where $dR=\prod_{i=1}^{N}d\textbf{r}_{i}$. Differentiating this function over time
and using the Schrodinger equation with the Hamiltonian (\ref{di BEC Hamiltonian}),
we derive the continuity equation (\ref{di BEC cont eq}). In the continuity equation the new quantity appears, it is the particle current or momentum density. We differentiate this quantity over time
and use the Schrodinger equation with the Hamiltonian (\ref{di BEC Hamiltonian}). In the results
we obtain the momentum balance equation. In this way we can find an evolution equation for any additive physical quantity, but we mast know its definition. The definition of new physical quantities appears at derivation of the QHD equation. Several quantities appears at derivation of the momentum balance equation (\ref{di BEC bal imp eq short}), one of them is the polarization presented below (\ref{di BEC def polarization}). Knowledge of the polarization definition gives possibility to derive the equation of polarization evolution (\ref{di BEC eq for pol current gen selfconsist
appr}) and so on.

In equations (\ref{di BEC bal imp eq short}), (\ref{di BEC field
eq}) polarization occurs in the form of
\begin{equation}\label{di BEC def polarization}P^{\alpha}(\textbf{r},t)=\int dR\sum_{i}\delta(\textbf{r}-\textbf{r}_{i})\psi^{*}(R,t)\hat{d}_{i}^{\alpha}\psi(R,t),\end{equation}
where $\textbf{r}_{i}$ is the coordinate of i-th particle, $dR=\prod_{p=1}^{N}d\textbf{r}_{p}$.

In the right-hand side of equation (\ref{di BEC eq for pol current gen
selfconsist appr}), where is the force-like field
$F^{\alpha\beta}(\textbf{r},t)$ which gives rise to evolution of
the polarization current $R^{\alpha\beta}(\textbf{r},t)$. In
general case, for $F^{\alpha\beta}(\textbf{r},t)$, we can write:
 \begin{equation}\label{di BEC }F^{\alpha\beta}(\textbf{r},t)=-\frac{1}{m}\partial_{\gamma}\Sigma^{\alpha\beta\gamma}(\textbf{r},t)+\frac{1}{m}D^{\alpha\gamma}(\textbf{r},t)\partial^{\beta}E^{\gamma}(\textbf{r},t).\end{equation}

The tensor
\begin{equation}\label{di BEC }D^{\alpha\beta}(\textbf{r},t)=\int dR\sum_{i}\delta(\textbf{r}-\textbf{r}_{i})d_{i}^{\alpha}d_{i}^{\beta}\psi^{*}(R,t)\psi(R,t),\end{equation}
occurs in the term that presents the dipole-dipole interaction and
an interaction of the dipole with an external electrical field. This value can be approximately
presented as
\begin{equation}\label{di BEC appr for D}D^{\alpha\beta}(\textbf{r},t)=\sigma\frac{P^{\alpha}(\textbf{r},t)P^{\beta}(\textbf{r},t)}{n(\textbf{r},t)},\end{equation}
that based on the reasons of dimensions.

The SRI causes the tensor $\Sigma^{\alpha\beta\gamma}(\textbf{r},t)$ to occur in the equation (\ref{di BEC eq for pol current gen selfconsist appr}). Taken at the first order by the interaction radius it has the form
$$\Sigma^{\alpha\beta\gamma}(\textbf{r},t)=-\frac{1}{2}\int dR\sum_{i,j\neq i}\delta(\textbf{r}-\textbf{R}_{ij})$$
\begin{equation}\label{di BEC short range in polariz}\times\frac{r_{ij}^{\beta}r_{ij}^{\gamma}}{r_{ij}}\frac{\partial U_{ij}}{\partial r_{ij}}\psi^{*}(R,t)\hat{d}_{i}^{\alpha}\psi(R,t).\end{equation}
Tensor $\Sigma^{\alpha\beta\gamma}(\textbf{r},t)$ describes the
influence of the SRI on evolution of polarization.
Formula (\ref{di BEC short range in polariz}) describes the SRI.

If we apply the procedure described in Ref. ~\cite{Andreev PRA08} to
calculate the quantum stress tensor, and neglect the
contribution of the thermal effects, when
$\Sigma^{\alpha\beta\gamma}(\textbf{r},t)$, for the BEC, takes the following form
\begin{equation}\label{di BEC 3ind-Sigma}\Sigma^{\alpha\beta\gamma}_{BEC}(\textbf{r},t)=-\frac{1}{2}\Upsilon\delta^{\beta\gamma}n(\textbf{r},t)P^{\alpha}(\textbf{r},t).\end{equation}
Formula (\ref{di BEC 3ind-Sigma}) is obtained for
particles located in state with the lowest energy, which can be
described by one particle wave function. However, this state may be the
product of strong interaction.
Tensor $\Sigma^{\alpha\beta\gamma}(\textbf{r},t)$, therefore,
like the quantum stress tensor
$\sigma^{\alpha\beta}(\textbf{r},t)$ in the momentum balance
equation (\ref{di BEC bal imp eq short}), depends on $\Upsilon$
at the first order by the interaction radius ~\cite{Andreev PRA08}.

\section{Appendix B:
Derivation of the non-linear Schr\"{o}dinger equation}

A derivation of GP equation was
performed for non-polarized particles from the QHD equations ~\cite{Andreev PRA08}, here we
present one for particles having electric dipole moment. The NLSE comes from the
continuity equation (\ref{di BEC cont eq}) and the Cauchy integral of the momentum
balance equation (\ref{di BEC bal imp eq short}). The Cauchy integral exists if the
velocity field can be expressed as
\begin{equation}\label{di BEC edde free cond}v^{\alpha}(\textbf{r},t)=\frac{1}{m}\partial^{\alpha}\theta(\textbf{r},t)\end{equation}
where $\theta(\textbf{r},t)$ is the velocity field potential.

Starting from the QHD equations we can derive an equation for
evolution of a model function defined in terms of hydrodynamic
variables. Thus, the macroscopic single-particle wave function may be
defined as
\begin{equation}\label{di BEC def psi func} \Phi(\textbf{r},t)=\sqrt{n(\textbf{r},t)}\exp\biggl(\frac{\imath}{\hbar}m\theta(\textbf{r},t)\biggr).\end{equation}

If we differentiate this function with respect to time and apply the
QHD equations when we obtain
$$\imath\hbar\partial_{t}\Phi(\textbf{r},t)=\Biggl(-\frac{\hbar^{2}}{2m}\nabla^{2}+\Gamma(\textbf{r},t)+g\mid\Phi(\textbf{r},t)\mid^{2}$$
$$-\int_{\textbf{r}_{0}}^{\textbf{r}}d\textbf{r}'\frac{1}{n(\textbf{r}',t)}P^{\beta}(\textbf{r}',t)\times$$
\begin{equation}\label{di BEC nlse int polariz}\times \nabla'\int
d\textbf{r}''G^{\beta\gamma}(\textbf{r}',\textbf{r}'')P^{\gamma}(\textbf{r}'',t)\Biggr)\Phi(\textbf{r},t),\end{equation}
here we use designation $\Gamma(\textbf{r},t)$ described by
formula
\begin{equation}\label{di BEC pressure term in kin appr} \Gamma(\textbf{r},t)=\int_{\textbf{r}_{0}}^{\textbf{r}} d\textbf{r}'\frac{\nabla p(\textbf{r}',t)}{n(\textbf{r}',t)}.\end{equation}
Quantity $\Gamma(\textbf{r},t)$ can be considered as chemical
potential $\mu(\textbf{r},t)$ and we use this below. The
equation (\ref{di BEC nlse int polariz}) has the form of a
NLSE. A NLSE with the DDI was obtained in Ref. ~\cite{Andreev PRB 11}, where the sign "-" was lost in front of integral in formulas (A6) and (A8).

The NLSE (\ref{di BEC nlse int polariz}) describes collective
properties of the many particles system. This follows from
the derivation of the NLSE and from the definition of the many-particle
wave function $\Phi(\textbf{r},t)$ (\ref{di BEC def psi func}).
Function $\Gamma(\textbf{r},t)$ (\ref{di BEC pressure term in kin
appr}) is the contribution of the kinetic pressure and does not
contain any interaction.

If we introduce a function $d^{\alpha}(\textbf{r},t)$ as $d^{\alpha}(\textbf{r},t)=P^{\alpha}(\textbf{r},t)/n(\textbf{r},t)$ and if also we suppose that $d^{\alpha}(\textbf{r},t)$ is a constant $d^{\alpha}$, when we obtain
$$\imath\hbar\partial_{t}\Phi(\textbf{r},t)=\Biggl(-\frac{\hbar^{2}}{2m}\nabla^{2}+\mu(\textbf{r},t)+g\mid\Phi(\textbf{r},t)\mid^{2}$$
\begin{equation}\label{di BEC nlse int polariz simple} -d^{\beta}d^{\gamma}\int
d\textbf{r}'G^{\beta\gamma}(\textbf{r},\textbf{r}')|\Phi(\textbf{r}',t)|^{2}\Biggr)\Phi(\textbf{r},t)\end{equation}
from equation (\ref{di BEC nlse int polariz}), we have used that $n(\textbf{r},t)=|\Phi(\textbf{r},t)|^{2}$.

At the temperature $T$ equal to zero in external electric field
$\textbf{E}_{0}$ all dipoles directed parallel to external field.
If $\textbf{E}_{0}=E_{0}\textbf{e}_{z}$ we have
$\textbf{d}=d_{0}\textbf{e}_{z}$. In this case in equation
(\ref{di BEC nlse int polariz simple}) only one component of Green
function $G^{\beta\gamma}$ of the DDI is nonzero, it is $G^{zz}$, which has following form
$G^{zz}=(1-3z^{2}/r^{2})/r^{3}+4\pi/3\cdot\delta(\textbf{r})$. we obtain the GP
equation (\ref{di BEC GP eq for introduction}) presented in
introduction at the absence of $\delta$ function
from (\ref{di BEC nlse int polariz simple}) and using relation $z^{2}/r^{2}=\cos^{2}\theta$.

The GP equation was obtained for the case of system of parallel
dipoles. We suppose it can be approximately used for analyze of
dipoles whose direction slowly changed in space.

\section{Appendix C}

We consider the QHD equations for the spinning particles to present it's difference from the QHD equations for particles having electric dipole moment.

Contribution of magnetization $\textbf{M}$ appears in the Euler equation (\ref{di BEC bal imp eq short}) instead
of polarization $\textbf{P}$. The method of QHD also allows to obtain equation of the magnetization
evolution
$$\partial_{t}M^{\alpha}(\textbf{r},t)+\nabla^{\beta}J^{\alpha\beta}_{M}(\textbf{r},t)$$
\begin{equation}\label{di BEC magn evol eq}=\frac{\gamma}{\hbar}\varepsilon^{\alpha\beta\gamma}M^{\beta}(\textbf{r},t)B^{\beta}(\textbf{r},t),\end{equation}
where $J_{M}^{\alpha\beta}$ arises. Vanishing  by thermal motion we
have $J_{M}^{\alpha\beta}=M^{\alpha}v^{\beta}$.

For obtaining of the QHD equations we started from the many-particle
Schrodinger equation
$$\imath\hbar\partial_{t}\psi_{s}(R,t)=\Biggl(\Biggl(\sum_{i}\biggl(\frac{p^{2}_{i}}{2m_{i}}-\gamma_{i}\hat{s}^{\alpha}_{i}B^{\alpha}_{i(ext)}\biggr) $$
\begin{equation}\label{di BEC spin gam}+\frac{1}{2}\sum_{i,j\neq i}\biggl(U_{ij}-\gamma_{i}\gamma_{j}G^{\alpha\beta}_{ij}\hat{s}^{\alpha}_{i}\hat{s}^{\beta}_{j}\biggr)\Biggr)\psi\Biggr)_{s}(R,t).\end{equation}
where we include the short-range and spin-spin interactions, and
action of an external magnetic field on spin. In the Schrodinger equation
(\ref{di BEC spin gam}) we use following designations:
$\gamma_{i}$ is the gyromagnetic ratio,
$\textbf{p}_{i}=-\imath\hbar\nabla_{i}$ is the operator of momentum,
$U_{ij}$ presents the short-range interaction, the Green function of
spin-spin interaction has form
$G^{\alpha\beta}_{ij}=4\pi\delta_{\alpha\beta}\delta(\textbf{r}_{12})+\nabla^{\alpha}_{1}\nabla^{\beta}_{1}(1/r_{12})$.

For spin matrixes $\hat{s}^{\alpha}_{i}$ the commutation relations
are
$$[\hat{s}^{\alpha}_{i},\hat{s}^{\beta}_{j}]=\imath\delta_{ij}\varepsilon^{\alpha\beta\gamma}\hat{s}^{\gamma}_{i}.$$
Thereby we consider Bose particles we present here the explicit
form of the spin matrixes $\hat{s}^{\alpha}_{i}$ for particles with
spin equal to 1:
$$\begin{array}{ccc} \hat{s}_{x}=\frac{1}{\sqrt{2}}\left(\begin{array}{ccc}0&
1&
0\\
1&
0&
1\\
0&
1&
0\\
\end{array}\right),&
\hat{s}_{y}=\frac{1}{\sqrt{2}}\left(\begin{array}{ccc}0& -\imath &
0\\
\imath &
0&
-\imath \\
0&
\imath &
0\\
\end{array}\right),&
\end{array}$$
$$\hat{s}_{z}=\left(\begin{array}{ccc}1&
0&
0\\
0& 0&
0\\
0& 0&
-1\\
\end{array}\right).$$

Derivation of the QHD equation from the Schrodinger equation we start
as usual from definition of the concentration of particles in vicinity of a
point $\textbf{r}$ of the physical space:
$$n(\textbf{r},t)=\sum_{s}\int
 dR\sum_{i}\delta(\textbf{r}-\textbf{r}_{i})\psi^{+}_{s}(R,t)\psi_{s}(R,t),$$
where $dR=\prod_{i=1}^{N}d\textbf{r}_{i}$. We obtain
equations analogous to (\ref{di BEC cont eq}) and (\ref{di BEC
bal imp eq short}), where magnetization
$\textbf{M}(\textbf{r},t)$ appears instead of polarization
$\textbf{P}(\textbf{r},t)$.

The magnetization arise in the form
$$M^{\alpha}(\textbf{r},t)=\sum_{s}\int dR\sum_{i}\delta(\textbf{r}-\textbf{r}_{i})\times$$
\begin{equation}\label{di BEC magnetization}\times\gamma_{i}\psi^{+}(R,t)_{s}(\hat{s}^{\alpha}_{i}\psi(R,t))_{s}.\end{equation}
Differentiating magnetization (\ref{di BEC magnetization}) with
respect to time and using Schrodinger equation (\ref{di BEC spin
gam}) we came to equation (\ref{di BEC magn evol eq}).

More details of obtaining of the QHD equations for spinning particles
are presented in Ref.s ~\cite{Andreev PRB 11}, ~\cite{Andreev arxiv MM}.

We can perform derivation of a NLSE for spinning particles, in the
result we obtain an equation analogous to (\ref{di BEC nlse
int polariz}). In the case parallel spins we have following
equation, where the difference between Green functions of the spin-spin and the
EDM interactions is accounted
$$\imath\hbar\partial_{t}\Phi(\textbf{r},t)=\Biggl(-\frac{\hbar^{2}}{2m}\nabla^{2}+\mu(\textbf{r},t)+g\mid\Phi(\textbf{r},t)\mid^{2}$$
\begin{equation}\label{di BEC nlse int magnet simple} +\mu_{0}^{2}\int
d\textbf{r}'\frac{1-3\cos^{2}\theta
'}{|\textbf{r}-\textbf{r}'|^{3}}|\Phi(\textbf{r}',t)|^{2}-\frac{8\pi}{3}\mu_{0}^{2}|\Phi(\textbf{r},t)|^{2}\Biggr)\Phi(\textbf{r},t).\end{equation}
Therefore, features of the spin-spin interaction give the additional term
in the GP equation for spinning particles, which is the last term in equation (\ref{di BEC nlse int magnet simple}). This term caused by the
long-range interaction, but it has form analogous to the SRI.

%\nocite{*}

\end{document}